\documentclass[useAMS,usenatbib]{mn2e}

\usepackage{newtxtext,newtxmath}

\usepackage[T1]{fontenc}
\usepackage{ae,aecompl}
\usepackage{natbib}
\def\fesc{\ifmmode f_{\rm esc} \else $f_{\rm esc}$\fi}

\def\apj{ApJ}
\def\apjs{ApJS}
\def\apjl{ApJL}
\def\aap{A\&A}
\def\aaps{A\&AS}

\def\mnras{MNRAS}

\def\pasp{PASP}
\def\nat{Nature}
\def\sci{Science}

\def\rmxaa{Revista Mexicana Astronomy and Astrophisics}
\def\araa{Ann.Rev.Astron.Astrophys.}

\usepackage{graphicx}	
\usepackage{amsmath}	
\usepackage{amssymb}	

\title[J1154$+$2443: a galaxy with 46 per cent LyC leakage]{J1154$+$2443: a low-redshift compact star-forming galaxy with a 46 per cent leakage of Lyman continuum photons}

\author[Y. I. Izotov et al.]{
Y. I. Izotov$^{1,2}$, 
D. Schaerer$^{3,4}$, 
G. Worseck$^{5,6}$, 
N. G. Guseva$^{1,2}$, 
T. X. Thuan$^{7}$, 
\newauthor 
~A. Verhamme$^{3}$, I. Orlitov\'a$^{8}$ \& K. J. Fricke$^{9,2}$
\\
$^{1}$Main Astronomical Observatory, Ukrainian National Academy of Sciences,
27 Zabolotnoho str., Kyiv 03143, Ukraine,\\ 
E-mail: izotov@mao.kiev.ua, guseva@mao.kiev.ua\\
$^{2}$Max-Planck-Institut f\"ur Radioastronomie, Auf dem H\"ugel 69, D-53121 
Bonn, Germany\\
$^{3}$Observatoire de Gen\`eve, Universit\'e de Gen\`eve, 
51 Ch. des Maillettes, 1290, Versoix, Switzerland,\\
E-mail: daniel.schaerer@unige.ch, anne.verhamme@unige.ch\\
$^{4}$IRAP/CNRS, 14, Av. E. Belin, 31400 Toulouse, France\\
$^{5}$Max-Planck-Institut f\"ur Astronomie, K\"onigstuhl 17, D-69117 Heidelberg, 
Germany\\
$^{6}$ Institut f\"ur Physik und Astronomie, Universit\"at Potsdam, Karl-Liebknecht-Str. 24/25, D-14476 Potsdam, Germany,\\
E-mail: gworseck@uni-potsdam.de\\
$^{7}$Astronomy Department, University of Virginia, P.O. Box 400325, 
Charlottesville, VA 22904-4325, USA,\\
E-mail: txt@virginia.edu\\
$^{8}$Astronomical Institute, Czech Academy of Sciences, Bo\v cn{\'\i} II 1401, 
141 00, Prague, Czech Republic,\\
E-mail: orlitova@asu.cas.cz\\
$^{9}$Institut f\"ur Astrophysik, G\"ottingen Universit\"at, 
Friedrich-Hund-Platz 1, D-37077 G\"ottingen, Germany,\\
E-mail: kfricke@gwdg.de}

\date{Accepted XXX. Received YYY; in original form ZZZ}

\pubyear{2017}

\begin{document}
\label{firstpage}
\pagerange{\pageref{firstpage}--\pageref{lastpage}}
\maketitle

\begin{abstract}
We report the detection of the Lyman continuum (LyC) radiation of
the compact star-forming galaxy (SFG) J1154$+$2443 observed with the Cosmic 
Origins Spectrograph (COS) onboard the {\sl Hubble Space Telescope}.
This galaxy, at a redshift of $z$ = 0.3690, is characterized by a high 
emission-line flux ratio O$_{32}$ = 
[O~{\sc iii}]$\lambda$5007/[O~{\sc ii}]$\lambda$3727 = 11.5.
The escape fraction of the LyC radiation $f_{\rm esc}$(LyC) 
in this galaxy is 46 per cent, the highest value found so far in 
low-redshift SFGs and one of the highest values found
in galaxies at any redshift. 
The narrow double-peaked Ly$\alpha$ emission line is detected in the spectrum of
J1154$+$2443 with a separation between the peaks $V_{\rm sep}$ 
of 199~km~s$^{-1}$, 
one of the lowest known for Ly$\alpha$-emitting galaxies, implying a high 
$f_{\rm esc}$(Ly$\alpha$). Comparing the extinction-corrected
Ly$\alpha$/H$\beta$ flux ratio with the case B value we find
$f_{\rm esc}$(Ly$\alpha$) = 98 per cent. 
Our observations, combined with previous detections in the
literature, reveal an increase of O$_{32}$ with increasing
$f_{\rm esc}$(LyC). We also find a tight anticorrelation 
between $f_{\rm esc}$(LyC) and $V_{\rm sep}$. 
The surface brightness profile derived from the COS acquisition image reveals 
a bright star-forming region in the centre and an exponential disc in the 
outskirts with a disc scale length $\alpha$ = 1.09 kpc. 
J1154$+$2443, compared to other known low-redshift LyC leakers, is characterized
by the lowest metallicity, 12+logO/H = 7.65$\pm$0.01, the lowest stellar mass 
$M_\star$ = 10$^{8.20}$~$M_\odot$, a similar star formation rate 
SFR = 18.9 $M_\odot$ yr$^{-1}$ and a high specific SFR of 
1.2$\times$10$^{-7}$ yr$^{-1}$.
\end{abstract}




\section{Introduction}\label{intro}

It is commonly accepted that star-forming galaxies were the main contributors
to the reionization of the Universe after the cosmic Dark Ages, when the baryonic 
matter was neutral. While \citet{Madau15} considered 
active galactic nuclei (AGN) as possible alternative sources of the 
ionization, more recent studies \citep*{H18,M18} show that the contribution of 
AGNs to the reionization of the Universe was small. However, the currently 
identified relatively bright star-forming galaxies (SFGs) are insufficient 
to fully ionize the Universe by redshift $z \sim 6$ \citep*{S01,C09,Iw09,R13}, and it is generally thought that numerous fainter low-mass SFGs, below the 
current detection limit of observations, are responsible for the 
bulk of the ionizing radiation at $z$ $\ga$ 6 \citep*{O09,WC09,M13,Y11,B15a}. 
Additionally, for galaxies to reionize the Universe, the escape fraction of 
their ionizing radiation has to be sufficiently high, of the order of 10--20 
per cent \citep[e.g.][]{O09,R13,D15,Robertson15,K16}. 

Generally, searches for Lyman continuum (LyC) leakers, both at high and low 
redshifts, have so far been difficult and largely unsuccessful.
Over the past, deep imaging studies at $z \sim 3$ have produced several 
candidate  LyC leaking galaxies
\citep[e.g.][]{S01,Iw09,N11,Mosta13}, whereas other teams have only obtained 
stringent upper limits \citep[e.g.][]{V10,B11}. 
Currently, the most reliable LyC leakers detected at high redshift 
are the objects {\em Ion2} \citep[$z=3.212$, ][]{Va15,B16} with a 
relative escape fraction $f^{\rm rel}_{\rm esc}$(LyC) = $0.64^{+1.1}_{-0.1}$, 
Q1549-C25 \citep[$z=3.212$, ][]{Sh16} with $f_{\rm esc}$(LyC) $>$ 0.51,
and A2218-Flanking \citep[$z\approx 2.5$, ][]{B17} with 
$f_{\rm esc}$(LyC) $> 0.28 - 0.57$.

Since direct observations of high-redshift galaxies are difficult because of 
their faintness, contamination by lower-redshift interlopers, and the increase 
of intergalactic medium (IGM) opacity \citep[e.g., ][]{V10,V12,Inoue14,Gr15}, 
it is important to identify and study local proxies of this galaxy population.
However, starburst galaxies at low redshifts are generally opaque to
their ionizing radiation \citep{L95,D01,Gr09}. This radiation was directly 
detected only in four low-redshift galaxies, with
small escape fractions $f_{\rm esc}$(LyC) of $\sim$1 -- 4.5 per cent. Two of these galaxies
were observed with the {\sl Hubble Space Telescope} ({\sl HST})/Cosmic 
Origins Spectrograph (COS) \citep{B14,L16}, one galaxy
with the {\sl Far Ultraviolet Spectroscopic Explorer} ({\sl FUSE}) \citep{L13}
and one galaxy with both the {\sl HST}/COS and {\sl FUSE} \citep{L13,L16}.

However, we note that according to the recent re-analysis of $f_{\rm esc}$(LyC) by
\citet{C17} in low-redshift galaxies observed with COS, only the galaxy of 
\citet{B14} is found to be a definite LyC
leaker. As for the galaxies discussed by \citet{L16}, Mrk~54 is not a LyC leaker
according to \citet{C17}, Tol~0440$-$381 and Tol~1247$-$232 are detected at 
the $\sim$~2$\sigma$ level, but these detections could be affected by residual 
geocoronal emission. Furthermore, the COS spectrum of Tol~1247$-$232 reveals 
a significantly lower $f_{\rm esc}$(LyC) than the earlier {\sl FUSE} data 
by \citet{L13} (0.4 per cent instead of 2.4 per cent). Haro 11 was observed
only with {\sl FUSE} \citep{L13} and has still to be confirmed with COS 
observations.

It was recently argued that low-mass compact galaxies at 
low redshifts $z < 1$ 
with very active star formation may be promising candidates for escaping 
ionizing radiation \citep{JO13,S15}. The subsample of these galaxies
in the redshift range $z$ $\sim$ 0.1 -- 0.3 is often referred to as ``Green Pea'' (GP) 
galaxies because of their green colour in the composite SDSS images \citep{Ca09}. On the other hand, 
\citet*{I11} have named all galaxies in the wider redshift range $\sim$ 0.0 -- 0.6
with an H$\beta$ emission-line luminosity $L$(H$\beta$) $\ga$ 10$^{40.5}$
erg s$^{-1}$ as Luminous Compact Galaxies (LCG).
The general characteristic of compact SFGs, including both GPs and LCGs,
is the presence of strong emission lines in the optical
spectra of their H~{\sc ii} regions, powered by numerous O-stars which produce
plenty of ionizing radiation.


  \begin{table}
 \caption{Coordinates, redshift, O$_{32}$ ratio and apparent magnitudes 
of J1154$+$2443
\label{tab1}}
\begin{tabular}{lr} \hline
R.A.(2000.0)           &11:54:48.85    \\
Dec.(2000.0)           &$+$24:43:33.03 \\
$z$                    &0.3690         \\
O$_{32}$                &11.5           \\
SDSS $u$ (mag)         &22.01$\pm$0.16 \\
SDSS $g$ (mag)         &21.77$\pm$0.05 \\
SDSS $r$ (mag)         &21.97$\pm$0.09 \\
SDSS $i$ (mag)         &21.99$\pm$0.13 \\
SDSS $z$ (mag)         &21.04$\pm$0.24 \\
{\sl GALEX} $FUV$ (mag)&22.10$\pm$0.53 \\
{\sl GALEX} $NUV$ (mag)&21.57$\pm$0.34 \\
\hline
\end{tabular}
  \end{table}


\citet{I15,I16c}, using SDSS data, selected $\sim$~15000 
compact SFGs at $z$ $<$ 1 and studied their properties. They found that in
general these galaxies are low-mass and low-metallicity galaxies. Their
stellar masses, star formation rates (SFR) and metallicities are similar
to those of high-redshift Lyman-alpha emitting (LAE) and Lyman-break 
galaxies (LBG) \citep{I15}. Many compact SFGs are characterised by 
high line ratios O$_{32}$ = 
[O~{\sc iii}]$\lambda$5007/[O~{\sc ii}]$\lambda$3727 $\ga$ 5, reaching
values of up to 60 in some galaxies \citep{S15}. Such high values
may indicate that H~{\sc ii} regions are density-bounded, allowing escape
of ionizing radiation to the IGM, as suggested e.g.\ by \citet{JO13} 
and \citet{NO14}. Indeed, \citet{I16,I16b}, using {\sl HST}/COS observations 
of five compact SFGs at redshift $z$ $\sim$ 0.3 and with O$_{32}$ = 5 -- 8, 
found that all these galaxies are leaking LyC radiation, with an  
escape fraction of 6 -- 13 per cent.

Based on the unique properties of compact SFGs and the success of our previous
observations \citep{I16,I16b}, we selected an additional sample of 6 more 
galaxies, covering a range of O$_{32}$ $>$ 10. J1154$+$2443 is the first of 
these galaxies that have been selected for spectroscopic observations with the 
{\sl HST}, in conjunction with the COS. Our aim is to detect
escaping ionizing radiation shortward of the Lyman continuum limit, at
rest wavelengths $\la$ 912\AA, and the Ly$\alpha$ emission line of this 
galaxy.

The results of these observations are presented in this paper.
The selection criteria are discussed in 
Section \ref{sec:select}. Extinction in the optical range and element abundances
are discussed in Section \ref{sec:ext}. The {\sl HST} observations 
and data reduction are described
in Section \ref{sec:obs}. We derive the surface brightness profile 
in the near-ultraviolet (NUV) range in Section \ref{sec:sbp}. 
Integrated characteristics of J1154$+$2443 are derived in 
Section \ref{sec:global}. 
Ly$\alpha$ emission is considered in 
Section \ref{sec:lya}. The Lyman continuum detection and the corresponding
escape fractions are presented in Section \ref{sec:lyc}. 
Finally, we summarize our findings in Section \ref{summary}.

\section{Selection criteria}\label{sec:select}

   The galaxy J1154$+$2443 was chosen from the sample of compact SFGs
selected from the SDSS Data Release 10 (DR10) \citep{A14}.
The following selection criteria were applied to construct the sample \citep{I15}:
1) the angular galaxy radius on the SDSS images $R_{50}$ $\leq$ 3 arcsec, 
where $R_{50}$ is the galaxy's Petrosian radius within which 50 per cent of the 
galaxy's flux in the SDSS $r$ band is contained;
2) a high equivalent width EW(H$\beta$) $\ga$ 50\AA\ of the H$\beta$ emission 
line in the SDSS spectrum; this ensures very recent star formation with 
an age of 2 -- 3 Myr and the presence of numerous hot O stars producing copious amounts of ionizing LyC 
radiation. Additionally, for the {\sl HST} observations, the following
criteria were applied:
1) a sufficiently high brightness in the far-ultraviolet (FUV) and a high 
enough redshift ($z \ga 0.3$) to allow direct LyC observations with the 
COS; and 2) a high extinction-corrected O$_{32} \ga 5$, which may indicate 
the presence of density-bounded H~{\sc ii} regions, i.e. escaping LyC radiation 
\citep*{JO13,NO14,I17}. 

J1154$+$2443 fulfils all the above criteria. In particular, its O$_{32}$ of
11.5 is the highest compared to all low-redshift LyC leakers observed so far
with the {\sl HST}/COS.

The coordinates, redshift, and O$_{32}$ ratio
of J1154$+$2443 and its SDSS and {\sl GALEX} apparent magnitudes are shown in 
Table \ref{tab1}.


  \begin{table}
  \caption{Extinction-corrected emission-line fluxes and equivalent widths 
in the SDSS spectrum
\label{tab2}}
  \begin{tabular}{lrrrr} \hline
Line &\multicolumn{1}{c}{$\lambda$} 
&\multicolumn{1}{c}{$I$$^{\rm a,b}$} &\multicolumn{1}{c}{$I$$^{\rm a,c}$} &\multicolumn{1}{c}{EW$_{\rm obs}$$^{\rm d}$} \\ \hline
Mg~{\sc ii}          &2796&  15.8$\pm$2.4&  11.9$\pm$1.8&   5 \\
Mg~{\sc ii}          &2803&   9.7$\pm$1.9&   7.3$\pm$1.5&   4 \\
$[$O~{\sc ii}$]$     &3727&  60.5$\pm$4.0&  50.4$\pm$3.5&  44 \\
H9                   &3836&  10.8$\pm$3.9&   9.3$\pm$3.4&   8 \\
$[$Ne~{\sc iii}$]$   &3869&  48.2$\pm$3.5&  42.8$\pm$3.1&  33 \\
H8+He~{\sc i}        &3889&  24.4$\pm$4.0&  21.4$\pm$3.5&  20 \\
H7+$[$Ne~{\sc iii}$]$&3969&  30.1$\pm$4.2&  26.7$\pm$2.7&  25 \\
H$\delta$            &4101&  30.3$\pm$4.1&  27.3$\pm$3.7&  26 \\
H$\gamma$            &4340&  50.7$\pm$3.9&  47.4$\pm$3.7&  74 \\
$[$O~{\sc iii}$]$    &4363&  17.8$\pm$2.4&  16.8$\pm$2.2&  19 \\
He~{\sc i}           &4471&   5.7$\pm$2.8&   5.5$\pm$2.3&  11 \\
H$\beta$             &4861& 100.0$\pm$5.1& 100.0$\pm$5.0& 220 \\
$[$O~{\sc iii}$]$    &4959& 194.2$\pm$7.5& 196.5$\pm$7.5& 466 \\
$[$O~{\sc iii}$]$    &5007& 568.0$\pm$16.& 577.5$\pm$16.&1121 \\
He~{\sc i}           &5876&  10.8$\pm$1.8&  11.7$\pm$1.9&  33 \\
H$\alpha$            &6563& 281.5$\pm$11.& 318.7$\pm$12.&1150 \\
$[$N~{\sc ii}$]$     &6583&   3.5$\pm$1.8&   3.9$\pm$2.0&  10 \\
$[$S~{\sc ii}$]$     &6717&   4.6$\pm$1.5&   5.3$\pm$1.7&  16 \\
$[$S~{\sc ii}$]$     &6731&   5.4$\pm$1.3&   6.2$\pm$1.5&  22 \\
He~{\sc i}           &7065&  10.6$\pm$2.1&  12.4$\pm$2.5&  65 \\
$C$(H$\beta$)$^{\rm e}$  &&\multicolumn{1}{c}{0.250}&\multicolumn{1}{c}{0.070}\\
$I$(H$\beta$)$^{\rm f}$       &&\multicolumn{1}{c}{13.8}&\multicolumn{1}{c}{9.1} \\ \hline
  \end{tabular}

$^{\rm a}$$I$ = 100$\times$$I$($\lambda$)/$I$(H$\beta$),
where $I$($\lambda$) and $I$(H$\beta$) are emission-line
fluxes, corrected for both the Milky Way and internal extinction.

$^{\rm b}$Corrected for internal extinction with $C$(H$\beta$) = 0.25 
derived from the observed Balmer decrement including H$\alpha$.

$^{\rm c}$Corrected for internal extinction with $C$(H$\beta$) = 0.07 
derived from the observed Balmer decrement excluding H$\alpha$.

\hbox{$^{\rm d}$Observed equivalent width in \AA.}

\hbox{$^{\rm e}$Internal extinction.}

\hbox{$^{\rm f}$in 10$^{-16}$ erg s$^{-1}$ cm$^{-2}$.}
  \end{table}


  \begin{table}
  \caption{Electron temperatures, electron number density and element 
abundances \label{tab3}}
  \begin{tabular}{lcc} \hline
Parameter &Value$^{\rm a}$&Value$^{\rm b}$ \\ \hline
$T_{\rm e}$ ($[$O {\sc iii}$]$), K      & 19100$\pm$1500& 18300$\pm$1400 \\
$T_{\rm e}$ ($[$O {\sc ii}$]$), K       & 15500$\pm$1100& 15300$\pm$1100 \\
$N_{\rm e}$ ($[$S {\sc ii}$]$), cm$^{-3}$&  1200$\pm$1000&  1200$\pm$1000 \\ \\
O$^+$/H$^+$$\times$10$^{5}$             &0.56$\pm$0.04&0.51$\pm$0.04  \\
O$^{2+}$/H$^+$$\times$10$^{5}$           &3.59$\pm$0.09&4.00$\pm$0.09 \\
O/H$\times$10$^{5}$                     &4.15$\pm$0.10&4.51$\pm$0.10 \\
12+log O/H                             &7.62$\pm$0.01&7.65$\pm$0.01  \\ \\
N$^+$/H$^+$$\times$10$^{6}$             &0.24$\pm$0.13&0.28$\pm$0.15  \\
ICF(N)$^{\rm c}$                &6.93&8.20  \\
N/H$\times$10$^{6}$                     &1.67$\pm$0.96&2.28$\pm$1.33 \\
log N/O                                &~$-$1.40$\pm$0.25~~\,&~$-$1.30$\pm$0.25~~\,  \\ \\
Ne$^{2+}$/H$^+$$\times$10$^{5}$          &0.68$\pm$0.06&0.67$\pm$0.06 \\
ICF(Ne)$^{\rm c}$               &1.06&1.04  \\
Ne/H$\times$10$^{5}$                    &0.72$\pm$0.06&0.70$\pm$0.06 \\
log Ne/O                               &~$-$0.76$\pm$0.04~~\, &~$-$0.81$\pm$0.04~~\, \\ \\
Mg$^{+}$/H$^+$$\times$10$^{7}$          &1.47$\pm$0.25&1.17$\pm$0.19 \\
ICF(Mg)$^{\rm c}$               &13.69&15.79  \\
Mg/H$\times$10$^{6}$                    &2.01$\pm$0.34&1.84$\pm$0.30 \\
log Mg/O                               &~$-$1.31$\pm$0.07~~\,&~$-$1.38$\pm$0.07~~\,  \\ \hline
\end{tabular}

\hbox{$^{\rm a}$For the internal extinction $C$(H$\beta$) = 0.25.}

\hbox{$^{\rm b}$For the internal extinction $C$(H$\beta$) = 0.07.}

\hbox{$^{\rm c}$Ionization correction factor.}
  \end{table}

\section{Extinction in the optical range and element abundances}\label{sec:ext}

The SDSS spectrum of J1154$+$2443 is characterized by a blue continuum
and strong nebular emission lines, indicating a young starburst
with an age of $\sim$ 2 -- 3 Myr. The observed decrement of several hydrogen Balmer 
emission lines is used to correct the line fluxes for reddening according
to \citet*{ITL94} and adopting the reddening law by \citet*{C89}
which is parameterized by the ratio of total-to-selective extinction
$R(V)=A(V)/E(B-V)$.
The extinction coefficient $C$(H$\beta$) derived 
from the hydrogen Balmer decrement corresponds to the extinction  
$A({\rm H}\beta) = 2.512 \times C({\rm H}\beta)$ at the H$\beta$ wavelength.
Then, adopting $R(V)$, the extinction $A(V)$ in the $V$ band is derived from the
relation $C$(H$\beta$)/$A(V)$ = $f$[$R(V)$] \citep{I16}.

The correction for reddening was done in two steps. First, the observed 
spectrum, uncorrected for redshift, was corrected for the Milky Way 
extinction with $A(V)_{\rm MW}$ from 
the NASA Extragalactic Database (NED)\footnote{NASA/IPAC Extragalactic
Database (NED) is operated by the Jet Propulsion Laboratory, California Institute of
Technology, under contract with the National Aeronautics and Space Administration.} and $R(V)_{\rm MW} = 3.1$.
Then, the rest-frame spectrum was corrected for the internal extinction 
of the galaxy, obtained from the hydrogen Balmer decrement.
One would expect to derive similar internal extinction whether the H$\alpha$ emission
line is included or not in the determination of $C$(H$\beta$). However, for 
J1154$+$2443 we find very different $C$(H$\beta$), of 0.250 with the
H$\alpha$ emission line included and of 0.070 with the H$\alpha$ emission line 
excluded. Adopting the high value $C$(H$\beta$) = 0.250 for reddening 
correction would result in an overcorrection 
of the H$\delta$/H$\beta$ and H$\gamma$/H$\beta$ ratios by $\sim$ 10 per cent
compared to the case B values (Table \ref{tab2}). 
This indicates that the H$\alpha$ emission line might be 
enhanced by some non-recombination processes. One of the possible mechanisms 
would be the collisional excitation of the H$\alpha$ line in the dense
H~{\sc ii} region of J1154$+$2443 \citep*[see e.g. ][]{I13}. 
Furthermore, the H$\alpha$ having an observed wavelength of 8986.8\AA\ and
full width at half maximum of 6\AA, it could 
be affected by residuals of the relatively strong night sky emission 
lines with wavelengths 8982.3\AA, 8985.7\AA\ and 8988.3\AA\ \citep{H03}.

The investigation of possible sources of the H$\alpha$ enhancement is
not the main concern of this paper. Therefore, we will adopt 
an empirical approach: we will consider both values of the
internal extinction coefficient, $C$(H$\beta$) = 0.25 and 0.07. We will then
compare the observational data in the UV and optical
ranges with our SED models to determine the most appropriate value 
of $C$(H$\beta$).

The extinction-corrected emission-line fluxes relative to the H$\beta$ 
emission line flux and the observed equivalent widths are shown 
in Table \ref{tab2} for both values of the internal extinction 
coefficient $C$(H$\beta$), 0.25 and 0.07.
The Table also gives $C$(H$\beta$) and the
H$\beta$ emission-line fluxes $I$(H$\beta$) corrected for both the Milky 
Way and internal extinctions derived in the way described above. 
It is seen that the extinction-corrected H$\delta$/H$\beta$ and 
H$\gamma$/H$\beta$ flux ratios with $C$(H$\beta$) = 0.07 are very close 
to the case B values, while the H$\alpha$/H$\beta$ flux ratio is $\sim$~15 per 
cent above the theoretical value. On the other hand, adopting 
$C$(H$\beta$) = 0.25 we derived case B H$\alpha$/H$\beta$ flux ratio, but
the H$\delta$/H$\beta$ and H$\gamma$/H$\beta$ flux ratios are $\sim$ 10 
per cent above the case B ratios.

We note that in the case of H {\sc ii} regions with non-negligible 
$f_{\rm esc}$(LyC), the recombination flux ratios of Balmer hydrogen lines
can deviate from the case B values. We check this possibility using 
the {\sc cloudy} models \citep{F98,F13} with different column densities
$N$(H~{\sc i}) and find that H$\alpha$/H$\beta$ and H$\gamma$/H$\beta$ 
respectively decrease and increase with decreasing $N$(H~{\sc i}), but
the effect is small, not exceeding 2 -- 3 per cent. Therefore, departures from
the case B cannot account for the observed hydrogen line ratios.

We use the emission line intensities (Table \ref{tab2}) and the direct 
$T_{\rm e}$-method to derive the electron temperature and electron number
density, and the ionic and total abundances of nitrogen, oxygen and neon in
the interstellar medium of J1154$+$2443 as described in \citet{I06}.
As for magnesium, the determination of its abundance is described in
\citet{G13}.

The derived temperatures and element abundances 
are shown in Table \ref{tab3}. The oxygen abundances of J1154$+$2443 are
7.65$\pm$0.01 with $C$(H$\beta$) = 0.07 and 7.62$\pm$0.01 with 
$C$(H$\beta$) = 0.25, lower than the oxygen abundances 
$\sim$ 7.8 -- 8.0 derived in known low-redshift LyC leakers by 
\citet{I16,I16b}. The ratios of other
element abundances to oxygen abundance are in the range obtained for
dwarf emission-line galaxies \citep[e.g. ][]{I06,G13}.


  \begin{table}
  \caption{{\sl HST}/COS observations \label{tab4}}
  \begin{tabular}{lcccc} \hline
\multicolumn{1}{c}{}&\multicolumn{1}{c}{}&\multicolumn{3}{c}{Exposure time (s)}\\ 
\multicolumn{1}{c}{Name}&\multicolumn{1}{c}{Date}&\multicolumn{3}{c}{(Central wavelength (\AA))} \\ 
    &    &MIRRORA&G140L&G160M \\ \hline
J1154$+$2443&2017-07-20& 1408    &  8691&  5636\\
            &          &         &(1105)&(1600)\\ \hline
\end{tabular}
  \end{table}

\section{{\sl HST}/COS observations and data 
reduction}\label{sec:obs}

\begin{figure}
\hbox{
\hspace{1.2cm}\includegraphics[angle=-90,width=0.80\linewidth]{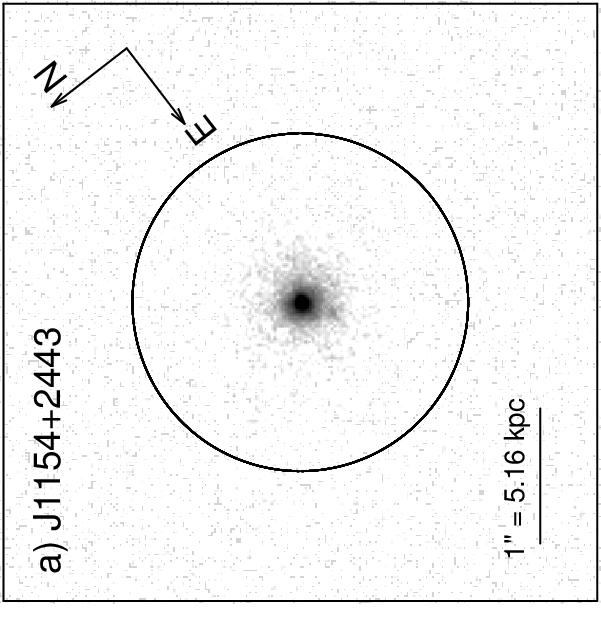}
}
\vspace{0.3cm}
\hbox{
\includegraphics[angle=-90,width=0.98\linewidth]{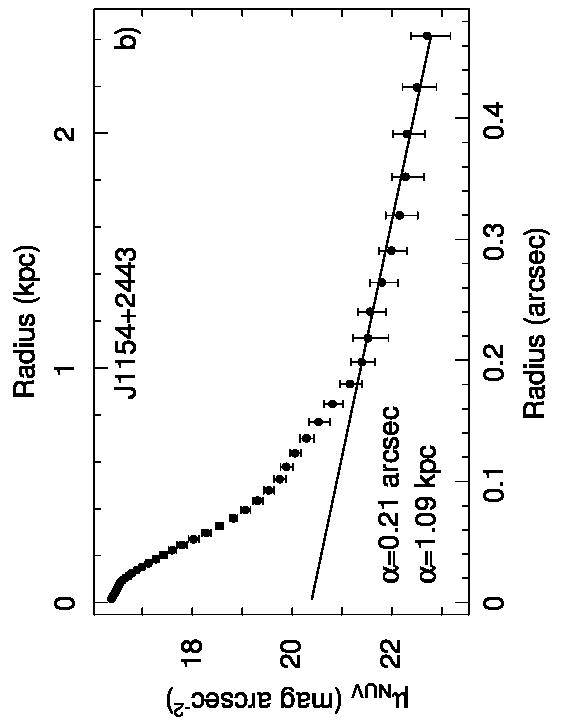}
}
\caption{{\bf a)} The {\sl HST}/COS NUV acquisition image of J1154$+$2443
in log surface brightness scale. The COS spectroscopic
aperture with a diameter of 2.5 arcsec is shown by a circle. The linear
scale is derived adopting an angular size distance of 1064~Mpc.
{\bf b)} Surface brightness profile of J1154$+$2443 derived from the NUV
acquisition image. The linear fit is shown for the range of
radii used for fitting.
\label{fig1}}
\end{figure}

The {\sl HST}/COS \citep{Gr12} observations of J1154$+$2443 were obtained in the
course of the program GO14635 (PI: Y. I. Izotov). Some details of the observations
are given in Table \ref{tab4}.
The galaxy was first acquired using COS near ultraviolet (NUV) images with the 
MIRRORA setting.
The galaxy region with the highest number of counts was automatically 
centered in the  2.5 arcsec diameter spectroscopic aperture 
(Fig. \ref{fig1}a). Although the galaxy shows some structure with an extended 
low-surface-brightness (LSB) component, it is very compact and is localized
in the central part of the spectroscopic aperture.

Spectra were obtained at COS Lifetime Position 3 with the gratings G140L and G160M,
applying four focal-plane offset positions in each observation to minimize fixed-pattern 
noise and to patch grid-wire shadows and other detector blemishes. 
The G140L grating was used in the 1105\AA\ setup
(wavelength range 1110--2150\AA, resolving power $R\simeq 1400$ at 1150\,\AA),
which includes the redshifted LyC emission.
The G160M spectrum was obtained in the 1600\AA\ setup
(wavelength range 1410--1773\AA, resolving power $R\simeq 17000$ at 1600\,\AA)
to resolve the redshifted Ly$\alpha$ $\lambda$1216\AA\ line and study its profile. 

The data were reduced with the {\sc calcos} pipeline v2.21 adapted to handle
spectra obtained at COS Lifetime Position 3, and custom software to improve
the background subtraction and co-addition for faint targets \citep{W16}.
We employed custom pulse height filtering to minimize the impact of detector dark current
by measuring the range of pulse heights with physical flux in the spectral trace.
For COS detector segment A we used pulse heights 1--12 containing $>99$ per cent of
the geocoronal Ly$\alpha$ flux. For segment B we estimated a pulse height range 3--15
from the spectral trace.
For boxcar extraction we used 25 (27) pixel wide rectangular apertures for the
G140L (G160M) spectra, preserving the spectrophotometry for compact sources
while minimizing the background.

The detector dark current in the chosen aperture was estimated from dark monitoring
exposures taken within $\pm 1$ month of the science observations in similar ambient
conditions \citep{W16}, and adopting the same pulse height cuts. We used the
pulse height distribution to select dark exposures obtained in similar ambient
conditions as the science data (see Appendix~\ref{sect:backgroundappendix} for details).
Low solar activity ensured that a sufficient number of dark exposures got selected,
resulting in a negligible statistical error ($\simeq 2$ per cent) of the dark current across the whole
wavelength range of both gratings. Tests with dark exposures revealed
that our dark subtraction procedure has negligible systematic error, and that the dark current
can be modelled accurately with dark exposures with similar pulse height distributions but
obtained at different orbit locations (Appendix~\ref{sect:backgroundappendix}).
The quasi-diffuse sky background was subtracted
using the {\sl GALEX} FUV map from \citet{M14}. Airglow contamination of N~{\sc i} 1200\AA\
and O~{\sc i} 1304\AA\ was eliminated by considering only data taken in orbital night
in the affected wavelength ranges, while N~{\sc i} 1134\AA\ and O~{\sc i}] 1356\AA\
line emission was negligible. For the G140L exposures we estimated
and subtracted scattered geocoronal Ly$\alpha$ emission following \citet{W16}.

We evaluated the accuracy of the estimated scattered light by comparing the fluxes
obtained in the total exposure time (8691\,s) and in orbital night (1797\,s). In the LyC of
the target the estimated total background error of 4--10 per cent does not significantly
affect our analysis.

\section{Surface brightness profile in the NUV range}\label{sec:sbp}

We use the COS NUV acquisition image of J1154$+$2443 to determine 
its surface brightess (SB) profile. The galaxy is sufficiently compact
with radius $\la$0.5 arcsec (Fig. \ref{fig1}a), therefore its image does not
suffer from vignetting. The image of J1154$+$2443 was reduced to the 
absolute scale by using the total {\sl GALEX} NUV magnitude (Table \ref{tab1}). 
Finally, the routine {\it ellipse} in {\sc iraf}\footnote{{\sc iraf} is distributed by the 
National Optical Astronomy Observatories, which are operated by the Association
of Universities for Research in Astronomy, Inc., under cooperative agreement 
with the National Science Foundation.}/{\sc stsdas}\footnote{{\sc stsdas} is 
a product of 
the Space Telescope Science Institute, which is operated by AURA for NASA.} 
was used to produce the SB profile. 
The SB profile shown in Fig. \ref{fig1}b consists of two parts,
one with a sharp SB increase in the central part, corresponding to 
the brightest star-forming region in the centre of the galaxy, and the second 
with a linear SB decrease (in magnitudes) of the extended low surface brightness component in the 
outward direction.

This linear decrease is characteristic of disc galaxies and can be described
by the relation
\begin{equation}
\mu (r)=\mu (0) + 1.086\times\frac{r}{\alpha},
\end{equation} 
where $\mu$ is the surface brightness, $r$ is the distance from the centre, 
and $\alpha$ is the disc scale length of the outer disc.

The derived scale length $\alpha$ $\sim$ 1.09 kpc (Fig. \ref{fig1}b) is very 
similar to scale lengths of other LyC leakers \citep{I16b}. The corresponding
surface density of star formation rate $\Sigma$ = SFR/($\pi \alpha^2$) is
5.1 M$_\odot$ yr$^{-1}$ kpc$^{-2}$.
The size of J1154$+$2443 in the UV can be directly compared with those
of galaxies at high redshifts. \citet{CL16}, \citet{PA17} and \citet{Bo17}
have shown that half-light radii of $z$ = 2 -- 8 galaxies are $\sim$ 0.1 -- 1 
kpc, which are similar to 
the J1154$+$2443 half-light radius of $\sim$ 0.2 kpc (Fig. \ref{fig1}b). 
Adopting the latter value we obtain much higher $\Sigma$ of 
$\sim$ 150 M$_\odot$ yr$^{-1}$ kpc$^{-2}$ which together with the SFR 
is a characteristic of the central part of the galaxy.


  \begin{table*} 
  \caption{Integrated parameters \label{tab5}}
  \begin{tabular}{lcccccccc} \hline
Name&12+logO/H$^{\rm a}$&  $D$$^{\rm b}$   &$M_g$$^{\rm a}$  &$M_{FUV}$$^{\rm a}$&log $M_\star$$^{\rm a,c}$ &SB age$^{\rm a}$&SFR$^{\rm a}$&$\alpha$$^{\rm d}$\\

    &         & (Mpc)  &(mag)  &(mag)&(log M$_\odot$)& (Myr)&(M$_\odot$ yr$^{-1}$)&(kpc) \\ \hline
J1154$+$2443&7.65$\pm$0.01&2064&$-$20.00&$-$20.12&8.20&2.6&18.9&1.09\\
\hline
  \end{tabular}

\hbox{$^{\rm a}$The internal extinction coefficient $C$(H$\beta$) = 0.07 is
adopted.}

\hbox{$^{\rm b}$Luminosity distance.}

\hbox{$^{\rm c}$$M_\star$ is the total mass of the young and old stellar populations.}

\hbox{$^{\rm d}$Exponential disc scale length.}

  \end{table*}

\section{Integrated characteristics of J1154$+$2443}\label{sec:global}

Using SDSS and {\sl GALEX} photometric and spectroscopic data, we derive in this 
Section integrated characteristics of J1154$+$2443 such as luminosities
and absolute magnitudes, SFR and stellar mass $M_\star$.

\subsection{Absolute magnitudes and the H$\beta$ luminosity}

The observed fluxes were transformed to absolute magnitudes and
luminosities adopting a luminosity distance derived with a cosmological 
calculator \citep[NED,][]{W06}, based on the cosmological 
parameters $H_0$=67.1 km s$^{-1}$Mpc$^{-1}$, $\Omega_\Lambda$=0.682, 
$\Omega_m$=0.318 \citep{P14}. We derived an extinction-corrected 
absolute SDSS AB $g$-band magnitude $M_g$ = $-$20.00 mag and a {\sl GALEX} 
FUV magnitude $M_{FUV}$ = $-$20.12 mag adopting $C$(H$\beta$) = 0.07
(Table \ref{tab5}). These magnitudes
are 0.5 -- 1.0 mag fainter than the respective values of the five
LyC leakers studied by \citet{I16,I16b}. This is expected because of the 
luminosity-metallicity relation for compact SFGs \citep{I15},
J1154$+$2443 having a lower metallicity compared to other LyC leakers.

The H$\beta$ luminosity $L$(H$\beta$) and SFR were obtained from the 
extinction-corrected H$\beta$ flux with $C$(H$\beta$) = 0.07. 
For J1154$+$2443 we derived $L$(H$\beta$) = 4.6$\times$10$^{41}$
erg s$^{-1}$. The SFR = 10.2 M$_\odot$ yr$^{-1}$ is determined from $L$(H$\beta$) 
using the relation by \citet{K98}. It should be increased 
by a factor 1/[1-$f_{\rm esc}$(LyC)] equal to 1.85, to take into account 
the escaping ionizing radiation
(see Sect.~\ref{sec:lyc}). This gives SFR = 18.9 M$_\odot$ yr$^{-1}$. 
It is at the low end of the SFR range 14 -- 50 $M_\odot$ yr$^{-1}$ 
derived by \citet{I16,I16b} for the five LyC leakers with higher metallicities
and absolute brightnesses. However, its specific star formation rate 
sSFR = SFR/$M_\star$ = 1.2$\times$10$^{-7}$ yr$^{-1}$ is the second highest in the
sample.

\begin{figure}
\begin{center}
\includegraphics[angle=-90,width=0.90\linewidth]{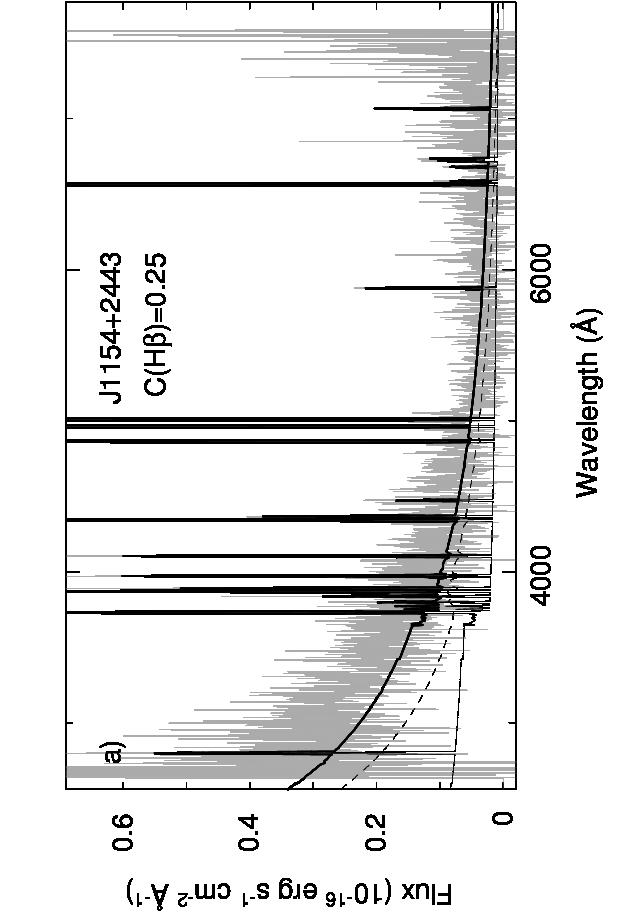}
\includegraphics[angle=-90,width=0.90\linewidth]{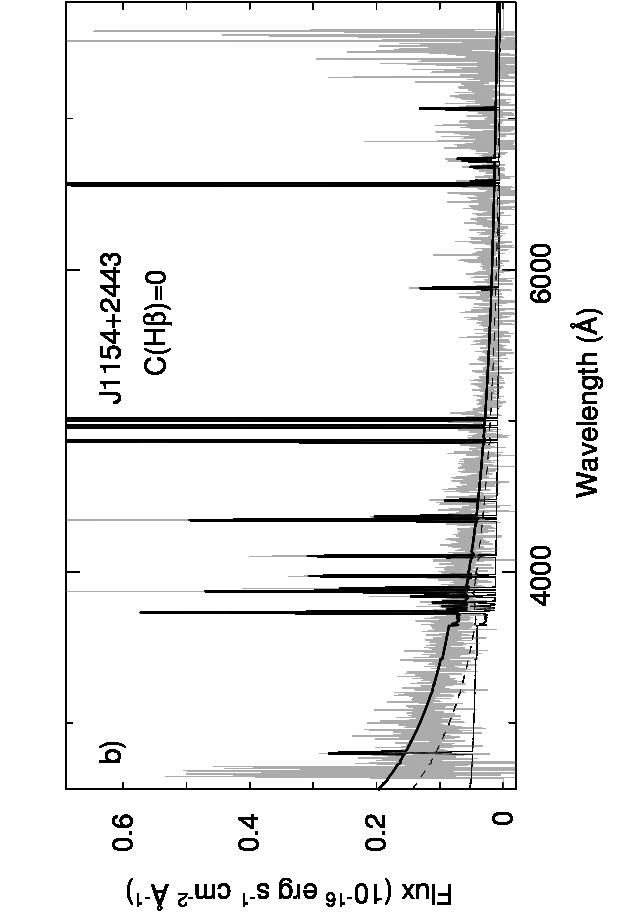}
\includegraphics[angle=-90,width=0.90\linewidth]{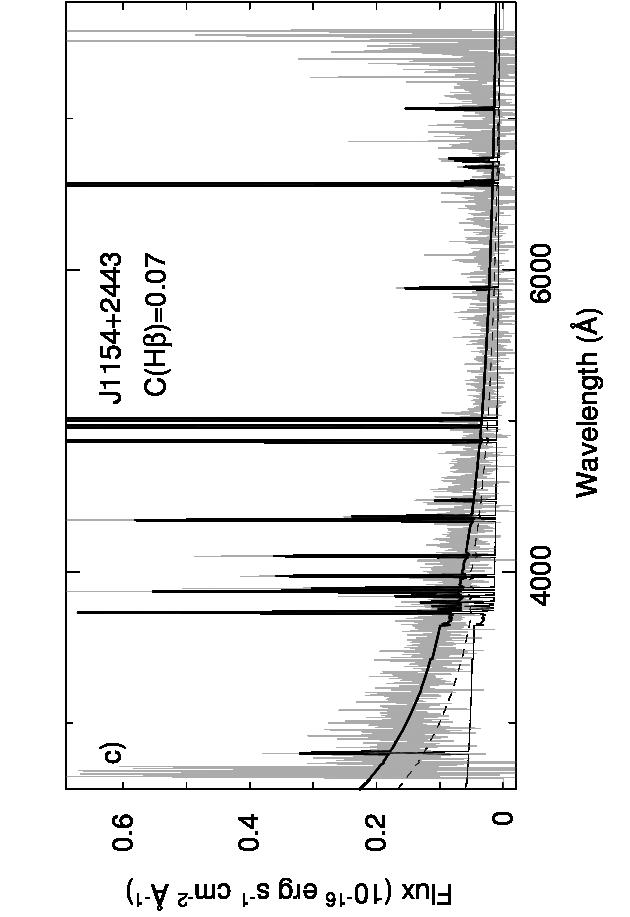}
\end{center}
\caption{SED fitting of the J1154$+$2443 optical spectrum. 
The rest-frame extinction-corrected SDSS spectrum is shown by the grey
line. The stellar, ionized gas, and total modelled SEDs are shown by dashed, 
thin solid and thick solid lines, respectively. {\bf a)} The case with the
internal extinction $C$(H$\beta$) = 0.25 derived including H$\alpha$ emission
line. {\bf b)} The case with $C$(H$\beta$) = 0. {\bf c)} The case with
$C$(H$\beta$) = 0.07 derived excluding H$\alpha$ emission
line.
\label{fig2}}
\end{figure}
%
\subsection{Stellar mass}

The stellar mass of J1154$+$2443 is derived from fitting the spectral energy 
distribution (SED) of the continuum in its SDSS spectrum in the wavelength range
3600--10300~\AA. We generally use the fitting method described e.g. in 
\citet{I16b}.
In J1154$+$2443, with a rest-frame H$\beta$ equivalent width EW(H$\beta$) 
$\sim$~160~\AA, nebular continuum is strong and should be accounted for in
SED modelling.
We note that the photometric data and UV spectra were not used in the SED 
fitting, but they are useful for checking the consistency of the SEDs derived 
from the optical spectra.

To fit the SEDs we carried out a series of Monte Carlo simulations.
A grid of instantaneous burst SEDs with a wide range of ages from 0 Myr to 
15 Gyr was calculated with {\sc starburst99} \citep{L99,L14} to derive
the SED of the galaxy stellar component. Input parameters for the grid
calculations included Padova stellar 
evolution tracks \citep{G00}, models of stellar 
atmospheres by \citet*{L97} and \citet*{S92}, and the stellar 
initial mass function of \citet{S55} with the upper and low stellar mass limits
of 100 M$_\odot$ and 0.1 M$_\odot$, respectively.
Then the stellar SED with any star-formation 
history can be obtained by integrating the instantaneous burst SEDs over
time with a specified time-varying SFR. 
Given the electron temperature $T_{\rm e}$ in the H~{\sc ii} region, we 
interpolated emissivities by \citet{A84} for the nebular continuum 
in the $T_{\rm e}$ range of 5000 -- 20000 K. 
Then the nebular continuum luminosity at any wavelength is derived from
the observed H$\beta$ luminosity. The fraction of the nebular continuum in
the total continuum near H$\beta$ is determined by the ratio of the observed
EW(H$\beta$) to the value of $\sim$ 1000\AA\ for the pure nebular emission.
The emission lines with fluxes 
measured in the SDSS spectrum were added to the total continuum.

The star-formation history is approximated assuming a
short burst with age $t_{\rm y}$ $<$ 10 Myr and a continuous star formation
for the older population with a constant SFR during the time interval between 
$t_i$ and $t_f$ ($t_f < t_i$ and zero age is now).
The contribution of each stellar population to the SED was parameterized by the
ratio $b$ = $M_{\rm y}$/$M_{\rm o}$, where $M_{\rm y}$ and $M_{\rm o}$ 
are the masses of the young and old stellar populations formed during the 
recent burst and the prior continuous star formation, respectively.

We calculated 10$^5$ Monte Carlo models to derive the stellar SED
by randomly varying $t_{\rm y}$, $t_i$, $t_f$, and $b$, while the fraction
of the nebular continuum is determined by the ratio of the observed 
EW(H$\beta$) to $\sim$ 1000\AA. 
We also calculate the equivalent widths EW(H$\beta$) and EW(H$\alpha$) for 
each model, integrating over the time 
luminosities of emission lines and adjacent continua. 
The best solution is required to fulfill the following conditions. First, only
models, in which both EW(H$\beta$) and EW(H$\alpha$) are in agreement with the 
observed values within 5 per cent, were selected. Second, the best modelled SED 
satisfying first two conditions was found from 
$\chi ^2$ minimization of the deviation between the modelled and the observed 
continuum.

\begin{figure}
\includegraphics[angle=-90,width=0.90\linewidth]{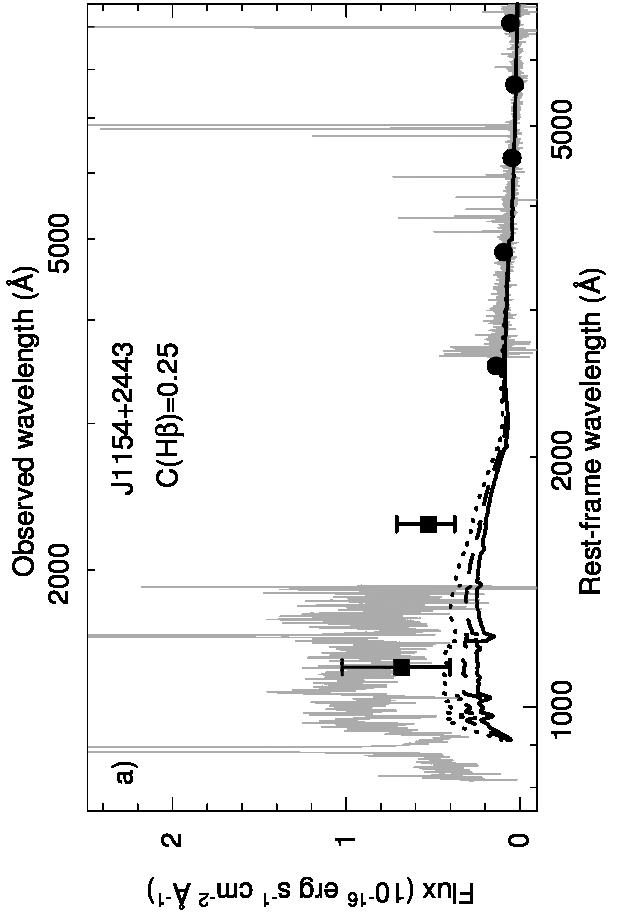}
\includegraphics[angle=-90,width=0.90\linewidth]{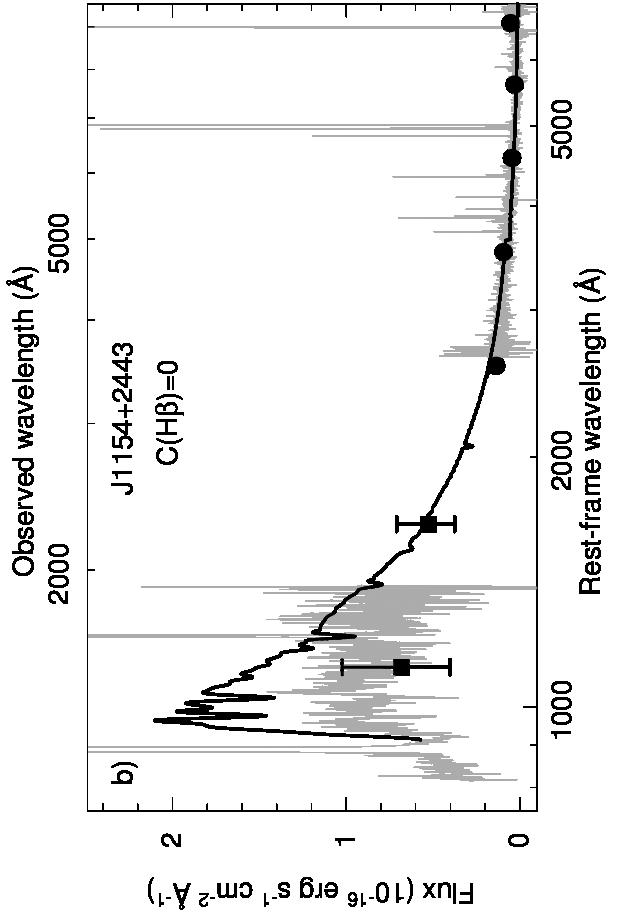}
\includegraphics[angle=-90,width=0.90\linewidth]{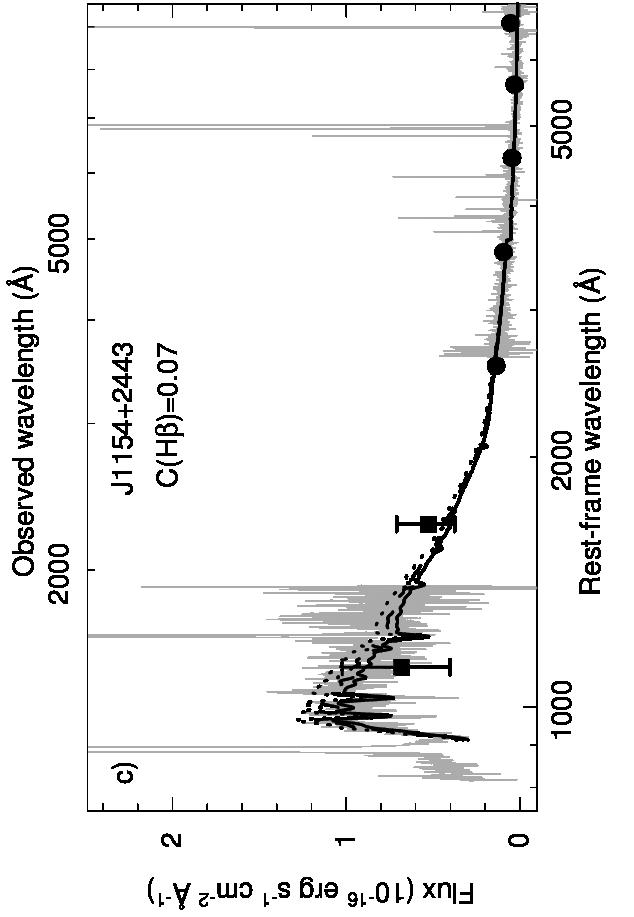}
\caption{A comparison of the observed UV and optical spectra, and
photometric data with the modelled SEDs.
The observed spectrum of J1154$+$2443 is shown by grey line. The total 
{\sl GALEX} and SDSS photometric fluxes are represented by filled squares and
filled circles, respectively. Modelled SEDs, which are reddened by the Milky 
Way extinction with $R(V)_{\rm MW}$ = 3.1 and internal extinction with 
$R(V)_{\rm int}$ = 3.1, 2.7, and 2.4 adopting the reddening law by \citet{C89} 
are shown by dotted, dashed and solid lines, respectively. 
{\bf a)} The case with the
internal extinction $C$(H$\beta$) = 0.25 derived including H$\alpha$ emission
line. {\bf b)} The case with $C$(H$\beta$) = 0. {\bf c)} The case with
$C$(H$\beta$) = 0.07 derived excluding H$\alpha$ emission
line.
\label{fig3}}
\end{figure}

We find from the best SED fit that the stellar mass 
$M_\star$~= 10$^{8.20}$~M$_\odot$ of J1154$+$2443. It is at the low end of the 
$M_\star$ range 10$^{8.22}$ -- 10$^{9.59}$~M$_\odot$ for other confirmed low-redshift 
LyC leakers \citep{I16,I16b}, and is typical to the mass of dwarf galaxies.

As for the starburst age $t_{\rm y}$, we note that it is mainly dependent on the 
H$\beta$ and H$\alpha$ equivalent
widths and is higher for lower EW(H$\beta$) and EW(H$\alpha$). 
The presence of escaping
ionizing radiation complicates the determination of the starburst age,
because this radiation does not produce H$\beta$ and H$\alpha$ emission. 
Therefore, EW(H$\beta$) and EW(H$\alpha$) are reduced in the case of large  
$f_{\rm esc}$(LyC),resulting in an overestimation of the starburst age.
To overcome this problem in the case of J1154$+$2443, during the SED fitting 
procedure, we have derived iteratively $f_{\rm esc}$(LyC) until
its value converges. We start by adopting an initial $f_{\rm esc}$(LyC) 
and deriving the best-fit SED by randomly varying 
$t_{\rm y}$, $t_i$, $t_f$, and $b$.
This model should also reproduce the observed EW(H$\beta$) and EW(H$\alpha$).
A new iterated value of $f_{\rm esc}$(LyC) is then derived as the ratio of the
observed LyC flux density to its intrinsic value as determined from the 
best-fit SED. Then the determination of the best-fit SED is repeated with the
new $f_{\rm esc}$(LyC). Only a few iterations are needed for
$f_{\rm esc}$(LyC) to converge, regardless of its initial value.
This iteration procedure adopting $C$(H$\beta$) = 0.07 results in a 
starburst age 
$t_{\rm y}$ = 2.6 Myr instead of 3.3 Myr, the value corresponding to $f_{\rm esc}$(LyC) = 0.
The determination of $f_{\rm esc}$(LyC) is further described in 
Sect. \ref{sec:lyc}.

The optical galaxy spectra with the overlaid stellar (dashed line), nebular
(thin solid line) and total stellar and nebular (thick solid line) best-fit
SEDs are shown in Fig. \ref{fig2} for $R(V)$ = 3.1 and three values of 
the internal extinction $C$(H$\beta$), 0.25 (Fig. \ref{fig2}a), 0 
(Fig. \ref{fig2}b), and 0.07 (Fig. \ref{fig2}c). It is seen that the 
observed spectrum shortward $\sim$3600\AA\ is not reproduced by the modelled 
SED if $C$(H$\beta$) = 0.25 is adopted (Fig. \ref{fig2}a). It is better
reproduced by the modelled SED adopting zero internal extinction
(Fig. \ref{fig2}b), but the best agreement is found if $C$(H$\beta$) = 0.07
is adopted (Fig. \ref{fig2}c). This conclusion is not changed if we adopt
$R(V)$ = 2.4 or 2.7. The stellar mass and starburst age derived 
from our fitting of the SDSS optical spectrum 
with $C$(H$\beta$) = 0.07 and $R(V)$ = 3.1 are presented in 
Table \ref{tab5}.

\subsection{Reddening law in the UV range}

We use the extrapolation to the UV range of the SED derived from the
optical SDSS spectrum. This is needed to verify what reddening law 
and internal extinction coefficient $C$(H$\beta$) are most
appropriate in the UV range of J1154$+$2443 in order to correct for extinction
observed fluxes in this range.
\citet{I16,I16b}, using the same approach, have shown that the correction for 
extinction in the optical range for LyC leakers is insensitive to variations 
of $R(V)$ because of low extinction. On the other hand they found 
that the observed spectra
of the LyC leakers in the entire UV + optical spectra are best fitted with
the reddening law by \citet{C89} adopting $R(V)_{\rm int}$ = 2.4 -- 2.7,
steeper than the canonical curve with $R(V)$ = 3.1, and resulting
in a larger UV extinction correction. 

The observed UV {\sl HST}/COS and optical SDSS spectra of J1154$+$2443 
are shown in Fig. \ref{fig3} by grey lines. Their fluxes are consistent
within the errors with the SDSS and {\sl GALEX}
photometric fluxes for the entire galaxy shown by the filled circles and 
filled squares, respectively. The consistency between the spectroscopic and 
photometric data in the optical range implies that almost all the galaxy 
emission is inside the spectroscopic aperture, and thus aperture corrections 
are not needed.

To compare with the observed spectrum, we reddened the intrinsic modelled 
SEDs in the optical range and their extrapolations in the UV range
for three values of the internal extinction coefficient $C$(H$\beta$),
0.25 (Fig. \ref{fig3}a), 0 (Fig. \ref{fig3}b) and 0.07 (Fig. \ref{fig3}c), 
adopting 
$R(V)_{\rm MW}$ = 3.1 for the Milky Way extinction, and $R(V)_{\rm int}$ = 3.1 
(dotted line), 2.7 (dashed line) and 2.4 (solid line) for 
the internal extinction. The Milky Way extinction was applied to the
SED redshifted to the observed wavelengths and the internal extinction to 
the spectrum at rest-frame wavelengths.
We adopted the reddening curve by \citet{C89} parameterized by the 
$R(V)_{\rm MW}$ and $R(V)_{\rm int}$, except for 
$\lambda\leq$ 1250\,\AA, where the reddening curve of \citet{M90} with the
respective $R(V)$'s is used.


  \begin{table*}
  \caption{Parameters for the H$\beta$ and Ly$\alpha$ emission 
lines \label{tab6}}
  \begin{tabular}{lccccccrcrcccccc} \hline
    &\multicolumn{3}{c}{}&&\multicolumn{3}{c}{H$\beta$}&\multicolumn{3}{c}{Ly$\alpha$}& \\
Name&$A(V)_{\rm MW}$$^{\rm a}$&$A(V)_{\rm int}$$^{\rm a}$&$A$(Ly$\alpha$)$^{\rm a}$&&$F$$^{\rm b}$&EW$^{\rm c}$&$L$$^{\rm d}$
&$F$$^{\rm b}$&EW$^{\rm c}$&$L$$^{\rm d}$&Ly$\alpha$/H$\beta$$^{\rm e}$&case B$^{\rm f}$&$f_{\rm esc}$(Ly$\alpha$) \\ \hline
J1154$+$2443&0.049&0.145&0.694&& 7.3&160&0.46& 148.9&133&14.38&31.3&31.9&0.98\\
\hline
  \end{tabular}

\hbox{$^{\rm a}$Extinction in mags. $A$(Ly$\alpha$) = $A$(Ly$\alpha$)$_{\rm MW}$ +
$A$(Ly$\alpha$)$_{\rm int}$, where $A$(Ly$\alpha$)$_{\rm MW}$ is derived for
$\lambda$=1216$\times$(1+$z$)\AA\ and $R(V)$=3.1,}

\hbox{ and $A$(Ly$\alpha$)$_{\rm int}$ 
is derived for $\lambda$=1216\AA\ and $R(V)$=2.4.}

\hbox{$^{\rm b}$Observed flux density in 10$^{-16}$ erg s$^{-1}$ cm$^{-2}$.}

\hbox{$^{\rm c}$Rest-frame equivalent width in \AA.}

\hbox{$^{\rm d}$Luminosity in 10$^{42}$ erg s$^{-1}$.}

\hbox{$^{\rm e}$Extinction-corrected flux ratio.}

\hbox{$^{\rm f}$Case B Ly$\alpha$/H$\beta$ flux ratio.}
  \end{table*}

\begin{figure}
\begin{center}
\includegraphics[angle=-90,width=0.98\linewidth]{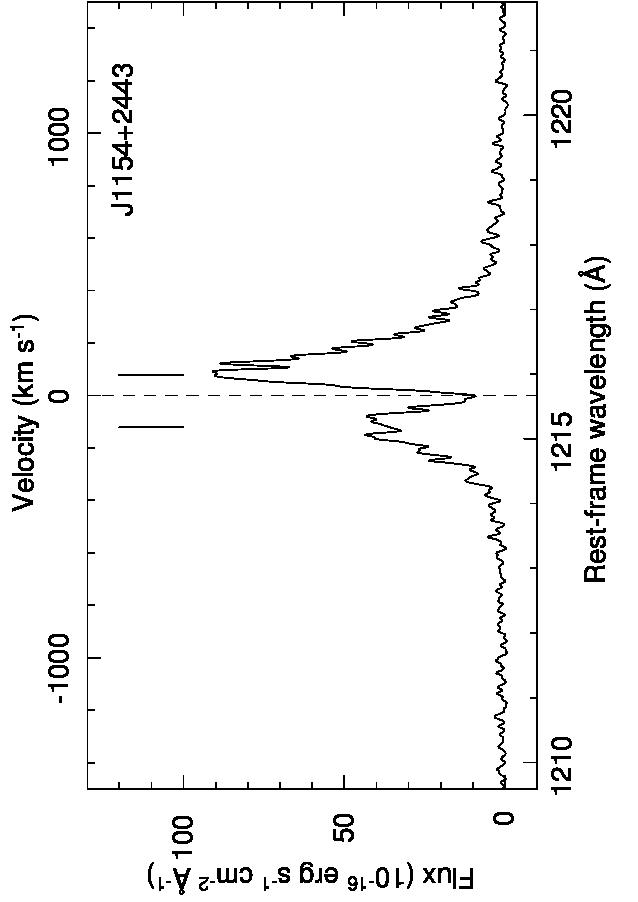}
\end{center}
\caption{The Ly$\alpha$ profile of J1154$+$2443. The centre of the line
is indicated by a vertical dashed line. The locations of the blue and red peaks
are shown by short solid vertical lines.
\label{fig4}}
\end{figure}

It is seen in Figs. \ref{fig3}a and \ref{fig3}b that the modelled
SEDs with $C$(H$\beta$) = 0.25 and 0 fail to reproduce the observed UV 
spectrum regardless of the $R(V)$ value, while the agreement is good 
with $C$(H$\beta$) = 0.07 (Fig. \ref{fig3}c). Therefore,  in the subsequent 
analysis, we will adopt the latter value of $C$(H$\beta$),
derived from the observed Balmer decrement excluding the H$\alpha$
emission line. We find that the observed UV spectrum of J1154$+$2443 
in  Fig. \ref{fig3}c is best fitted by the 
SED reddened with $R(V)_{\rm int}$ = 2.4, in close agreement with conclusions 
made by \citet{I16,I16b} for five LyC leakers, while the SEDs reddened with
higher $R(V)_{\rm int}$ lie above the observed UV spectrum, requiring
a higher extinction in the UV range than that derived from the Balmer
decrement. \citet{I16,I16b}
also considered reddening laws by \citet*{C94} and \citet{C00} and found that
the SEDs reddened with this law do not fit the observed UV spectra of LyC
leakers, requiring a higher extinction in the UV range than that by
\citet{C89} with $R(V)_{\rm int}$ = 3.1. For subsequent discussion 
of the J1154$+$2443 UV spectrum, we adopt $R(V)_{\rm int}$ = 2.4.

\section{Ly$\alpha$ emission}\label{sec:lya}

Strong Ly$\alpha$ $\lambda$1216\,\AA\ emission-line is detected in the 
medium-resolution spectrum of J1154$+$2443 (Fig. \ref{fig4}). 
The profile shows two 
peaks at $-$121 km s$^{-1}$ and $+$78 km s$^{-1}$, labelled by two short 
vertical lines in the Figure. While the shape of the Ly$\alpha$ profile 
in J1154$+$2443 is similar to that observed in known LyC leakers \citep{V17}
and in some other GP galaxies \citep{JO14,H15,Y17}, the separation between the
blue and red peaks of 199 km s$^{-1}$ is the lowest among all low-redshift
LyC leakers and other GP galaxies observed so far with the {\sl HST}/COS.
According to the radiation transfer models of \citet{V15}, such a low
separation indicates a very low column density of the neutral gas 
in J1154$+$2443 and thus a high escape fraction
of the Ly$\alpha$ emission. 
We find that the rest-frame equivalent width EW(Ly$\alpha$) of the 
Ly$\alpha$ line equal to 133\AA\ and the extinction-corrected luminosity 
$L$(Ly$\alpha$) (Table \ref{tab6}) 
are among the highest for the low-redshift LyC leakers \citep{I16,I16b}.

\begin{figure*}
\hbox{
\includegraphics[angle=-90,width=0.48\linewidth]{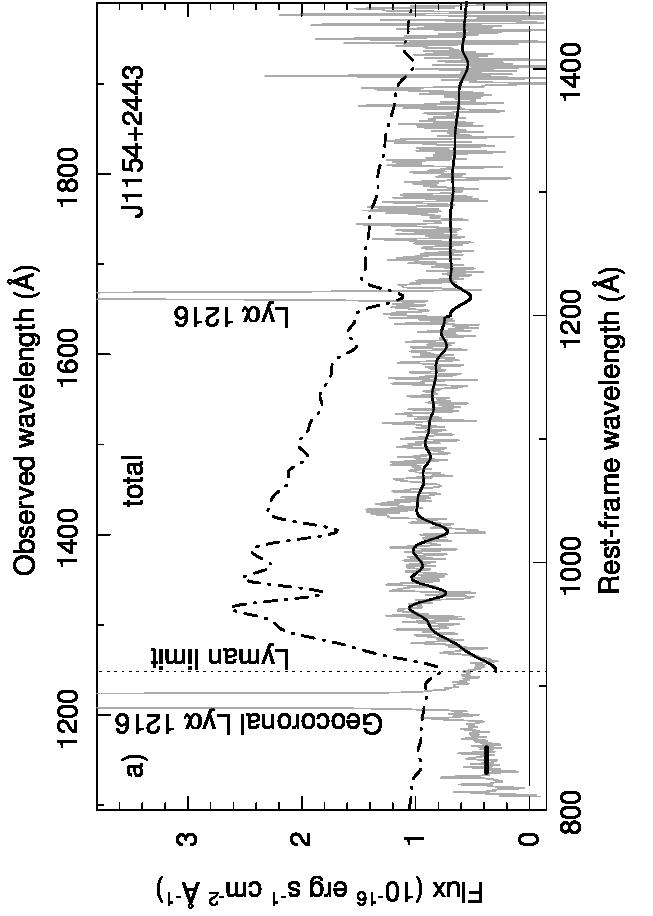}
\hspace{0.4cm}\includegraphics[angle=-90,width=0.48\linewidth]{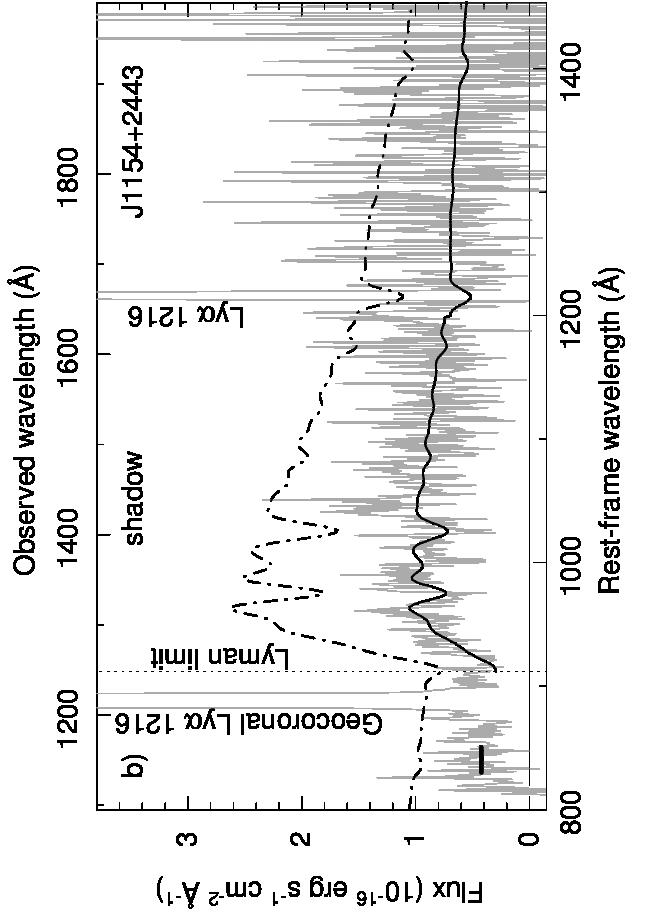}
}
\hbox{
\includegraphics[angle=-90,width=0.48\linewidth]{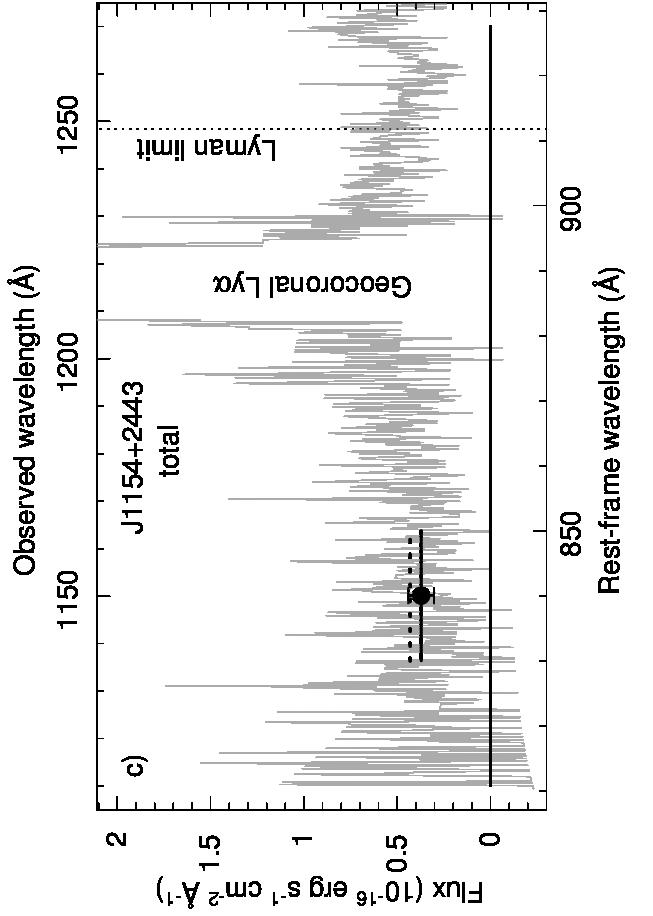}
\hspace{0.4cm}\includegraphics[angle=-90,width=0.48\linewidth]{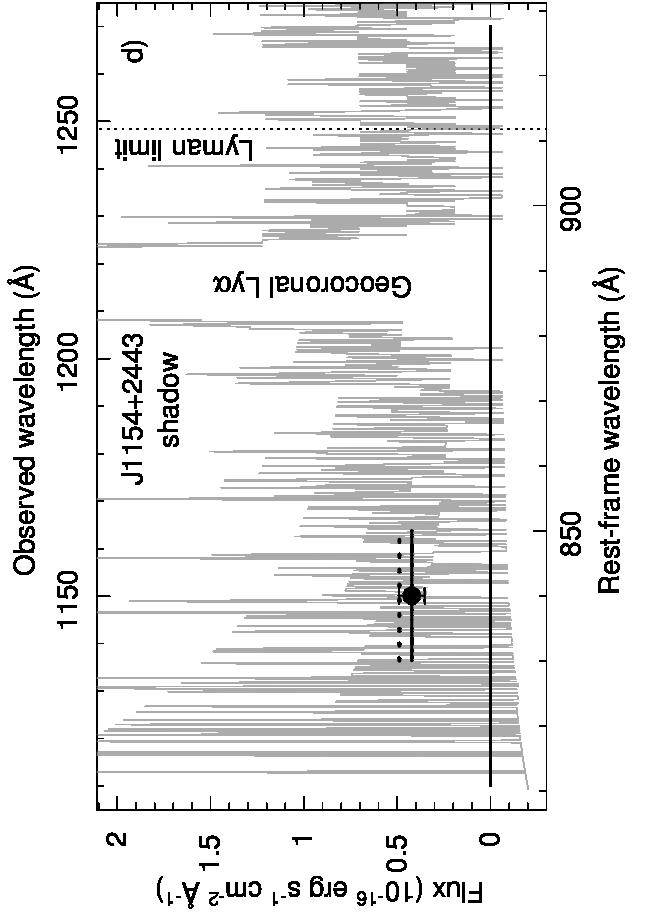}
}
\caption{{\bf a)} COS G140L spectrum of J1154$+$2443 obtained with an
exposure of 8691~s and smoothed by 12-pixel binning. On top of the 
observed spectrum (grey line) is superposed the modelled SED, reddened
by both the Milky Way and internal extinctions (solid line). The unreddened 
intrinsic SED is shown by the dash-dotted line. The thick and short solid 
horizontal line at the
rest-frame $\lambda$ = 830 -- 850\AA\ shows the level of the observed LyC flux 
density. 
Zero flux is represented by a black solid horizontal line while 
the Lyman limit is indicated by the vertical dotted line.
{\bf b)} Same as in {\bf a)} but the spectrum is taken only in orbital shadow 
with an exposure time of 1797 s.
{\bf c)} Expanded COS spectrum with the total exposure time of 8691 s
showing the Lyman continuum. The spectrum is smoothed by 3-pixel binning.
The average value 
of the observed LyC flux density with the 3$\sigma$ error bar is shown by the
filled circle. The solid horizontal line indicates the observed mean LyC flux 
density and the wavelength range of 830 -- 850\AA\ used for averaging, while 
the dotted horizontal
line represents the LyC level after correction for the Milky Way extinction. 
Zero flux is also shown by a horizontal line. {\bf d)} Same as in {\bf c)}
but the spectrum taken in orbital shadow is shown.
\label{fig5}}
\end{figure*}


  \begin{table*}
  \caption{LyC escape fraction \label{tab7}}
\begin{tabular}{lccccl} \hline
 &\multicolumn{3}{c}{Monochromatic flux density$^{\rm a}$}& \multicolumn{2}{c}{LyC escape fraction} \\
Name&$I_{\rm obs}$(840)$^{\rm b}$&$I_{\rm esc}$(840)$^{\rm c}$&$I_{\rm mod}$(840)$^{\rm d}$&\multicolumn{1}{c}{$f_{\rm esc}$$^{\rm e}$}&$f_{\rm esc}$$^{\rm f}$ \\
\hline
J1154$+$2443&3.8$\pm$0.2&4.4$\pm$0.2&9.6&0.46$\pm$0.02&0.47 \\
\hline
  \end{tabular}

\hbox{$^{\rm a}$in 10$^{-17}$ erg s$^{-1}$cm$^{-2}$\AA$^{-1}$.}

\hbox{$^{\rm b}$Observed flux density in the wavelength range 830 -- 850\AA.}

\hbox{$^{\rm c}$Flux density corrected for the Milky Way extinction.}

\hbox{$^{\rm d}$Intrinsic flux density derived from the modelled SED.}

\hbox{$^{\rm e}$$I_{\rm esc}$/$I_{\rm mod}$(840) (first method, Eq. \ref{eq:fesc}).}

\hbox{$^{\rm f}$$I_{\rm esc}$/$I_{\rm mod}$(840) (second method, Eqs. \ref{eq:i900} and
\ref{eq:fesc1}).}

  \end{table*}

Comparing the extinction-corrected Ly$\alpha$/H$\beta$ flux ratio
of 31.3 (Table \ref{tab6}) and the 
case B flux ratio of 31.9 obtained with the {\sc cloudy} code \citep{F98,F13}
for an electron temperature $T_{\rm e}$ = 18300K and an electron number density 
$N_{\rm e}$ = 1000 cm$^{-3}$, we find that the Ly$\alpha$ escape fraction is 
 $f_{\rm esc}$(Ly$\alpha$) = 98 per cent, the highest known so far for LyC leakers
and other GP galaxies \citep{H15,I16,I16b,Y17}. 

\section{Escaping Lyman continuum radiation}\label{sec:lyc}

Our primary aim in this paper is to examine the COS spectrum 
of J1154$+$2443 obtained with 
the G140L grating and to derive the escape fraction of ionizing radiation.

\subsection{Low-resolution G140L spectrum of J1154$+$2443}

The observed G140L spectrum with a total exposure of 8691~s is shown in 
Fig. \ref{fig5}a together with the 
predicted intrinsic SED obtained from fitting the observed optical SDSS spectrum
(see Section \ref{sec:global}). The strong line 
at the observed wavelength of 1216\AA\ is the residual of the
geocoronal Ly$\alpha$ emission, while the second brightest line
labelled ``Ly$\alpha$ 1216'' is the Ly$\alpha$ line of the galaxy. 
It is seen that the flux shortward the Lyman limit at the rest-frame
wavelength 912\AA\ indicated by a dotted vertical line is far above the zero 
level. 

There is a slight increase of the flux in the rest-frame wavelength range 
855 -- 912\AA\ centered at the geocoronal Ly$\alpha$ emission line. 
The first suggestion is that this 
increase is  due to the geocoronal Ly$\alpha$ scattered light.
However, we argued in Section \ref{sec:obs} that the scattered light might be 
not significant.
To validate this point we use the COS spectrum of a quasar at $z$ = 3.9
that shows highly saturated He {\sc ii} absorption with the rest-frame
wavelength of $\sim$ 240\AA\ and therefore with a very little source flux 
which is redshifted to the wavelength range near the geocoronal Ly$\alpha$ line.
In this case, almost all emission would be attributed to the scattered 
geocoronal Ly$\alpha$ emission line. However, we find that geocoronal flux is 
small at $\lambda$ $>$ 1230\AA\ and at 1180 -- 1200\AA.
This indirect test shows that the scattered geocoronal Ly$\alpha$ 
emission is insufficient to explain the flux increase around the geocoronal 
Ly$\alpha$ emission line. 

To further explore whether scattered geocoronal
Ly$\alpha$ emission is important, we compare in Fig. \ref{fig5}b the spectrum 
during orbital shadow with an exposure 1797 s 
when the flux of the geocoronal line is 
decreased by a factor of $\sim$ 20 with respect to the total spectrum (Fig. \ref{fig5}a). 
It is seen that the presence of the
flux increase around the geocoronal Ly$\alpha$ line is not so evident,
although we cannot totally exclude the contribution of the scattered 
geocoronal Ly$\alpha$ light in the case of J1154$+$2443. 
However, the signal-to-noise ratio in Fig. \ref{fig5}b is not sufficient 
to make more definite conclusions. Therefore, to be safe, we exclude the
wavelength range 855 -- 912\AA\ and use the range
830 -- 850\AA\ (corresponding to the observed wavelength range 
1136 -- 1164\AA) for the LyC flux determination. 
The average flux density at these wavelengths, attributed to the escaping LyC 
emission of J1154$+$2443, is indicated in Figs. \ref{fig5}a and \ref{fig5}b by 
the short horizontal solid lines.

This is illustrated more clearly in 
Figs. \ref{fig5}c and \ref{fig5}d, which show a blow-up of the LyC spectral 
region. With the flux density of 
$(3.8\pm0.2) \times 10^{-17}$ erg s$^{-1}$ cm$^{-2}$ \AA$^{-1}$
averaged over the rest-frame spectral range 830 -- 850\AA, 
the Lyman continuum is detected at the $\sim$ 18$\sigma$ level
in the total exposure spectrum.

\begin{figure*}
\includegraphics[angle=-90,width=0.48\linewidth]{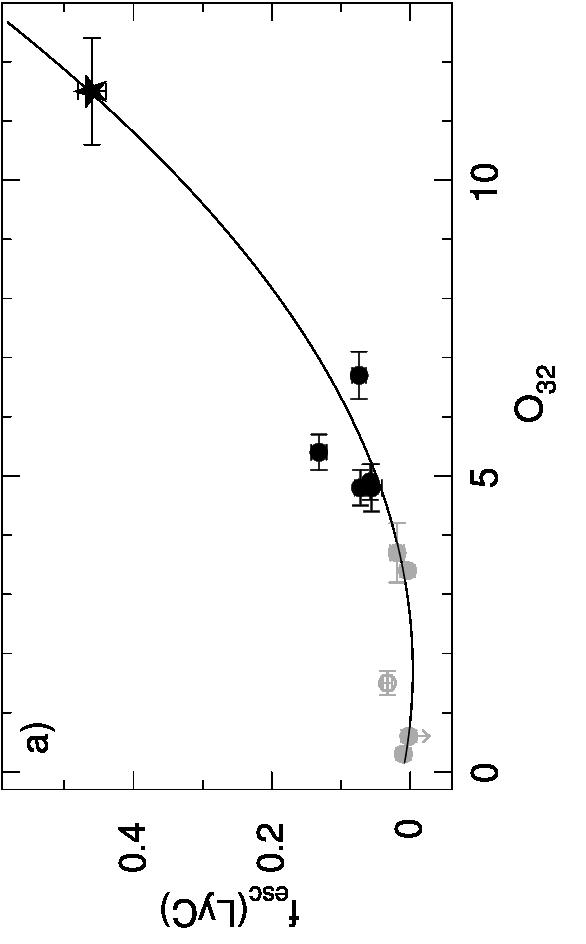}
\includegraphics[angle=-90,width=0.48\linewidth]{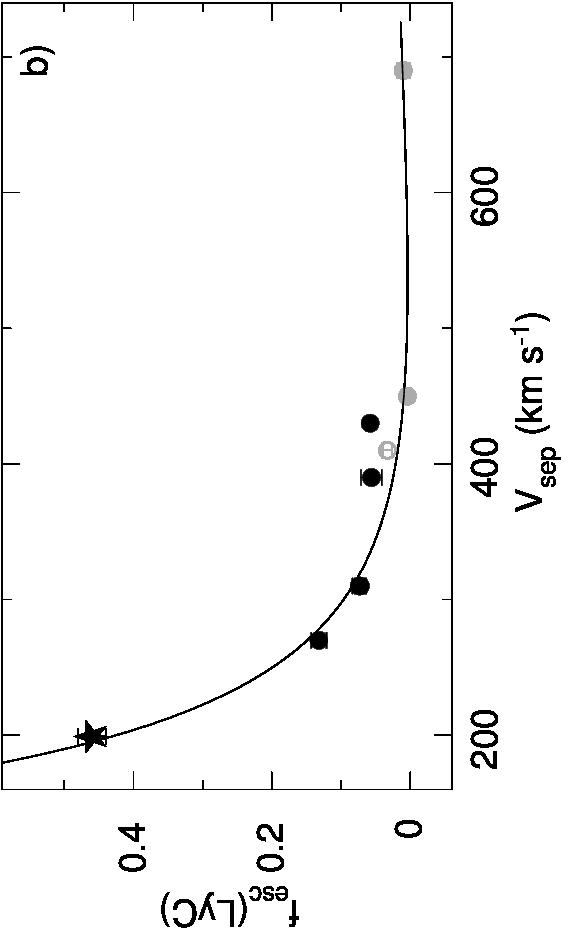}
\caption{{\bf a)} Relation between the Lyman continuum escape fraction
$f_{\rm esc}$(LyC) and the O$_{32}$ = 
[O~{\sc iii}]$\lambda$5007/[O~{\sc ii}]$\lambda$3727 
emission-line ratio for low-redshift LyC leaking galaxies observed with the
{\sl FUSE} (open circle) and {\sl HST}/COS (filled symbols). Compact SFGs from
\citet{I16,I16b} are shown by black filled circles. The galaxy from
\citet{B14} and the galaxies from \citet{L16} with $f_{\rm esc}$(LyC) derived 
by \citet{C17} are represented by grey filled circles. Haro 11 \citep{L13} and
J1154$+$2443 are shown by the open circle and filled star, respectively. 
{\bf b)} Relation between 
$f_{\rm esc}$(LyC) and a separation $V_{\rm sep}$ between the blue and red peaks
of the Ly$\alpha$ emission line for low-redshift LyC leaking galaxies
observed with the {\sl FUSE} and {\sl HST}/COS.
$V_{\rm sep}$ for J1154$+$2443 is from this paper and for other galaxies are
from \citet{V17}. Symbols are the same as in {\bf a)}. Solid lines in both 
panels are most likelihood regressions.
\label{fig6}}
\end{figure*}

The observed LyC emission should be corrected for extinction from the Milky Way 
before the determination of the LyC escape fraction. The average LyC flux 
densities of $4.4 \times 10^{-17}$ erg~s$^{-1}$~cm$^{-2}$~\AA$^{-1}$
and $4.8 \times 10^{-17}$ erg~s$^{-1}$~cm$^{-2}$~\AA$^{-1}$,
corrected for the Milky Way extinction, are shown respectively 
in Figs. \ref{fig5}c and \ref{fig5}d
by the dotted horizontal lines. The observed and corrected flux densities 
derived from the total exposure which we adopt, 
are reported in Table \ref{tab7}.

\subsection{LyC escape fraction of J1154$+$2443}

To derive the escape fraction of Lyman continuum photons, $f_{\rm esc}$(LyC), 
\citet{I16,I16b} used the modelled flux density $I_{\rm mod}(900)$ of 
the Lyman continuum emission at 900 \AA\ and compared it to the observed flux
density (after correction for extinction in the Milky Way).
They proposed two methods to derive $I_{\rm mod}(900)$. 
The first method is based on the SED fit to the observed SDSS spectrum in the
optical range described in Sect. \ref{sec:global}. The extrapolation of this
fit predicts the UV emission, including the Lyman continuum.
The modelled UV flux densities $I_{\rm mod}$($\lambda$) are shown 
by the dash-dotted line in Fig. \ref{fig5}a. The reddened modelled UV SED is 
represented by the solid line. This SED is the same as the SED shown by the solid 
line in Figs. \ref{fig2} -- \ref{fig3} for $R(V)_{\rm MW}$ = 3.1, $A(V)_{\rm MW}$ 
= 0.049 mag 
from the NED, $A(V)_{\rm int}$ = 0.145 mag as derived from the Balmer decrement,
and $R(V)$ = 2.4 (see details in Sect. \ref{sec:global}). 
It is seen that the reddened SED reproduces very well 
the observed spectrum (grey line) for rest-frame wavelengths $>$ 912\AA.

The second method of the $I_{\rm mod}$ determination uses the fact that the 
intensities of hydrogen recombination lines are approximately proportional to 
the number of ionizing photons emitted per unit time, $N$(Lyc).

Both methods use the extinction-corrected flux density $I$(H$\beta$) 
and rest-frame 
equivalent width EW(H$\beta$) of the H$\beta$ emission line, both derived
from observations.
Since the observed LyC flux density is measured in the wavelength range 
830 -- 850\AA, because the flux density at 900\AA\ might be affected by 
geocoronal Ly$\alpha$ emission, we adopt $\lambda$ = 840\AA.
Then the escape fraction $f_{\rm esc}$(LyC) is derived from
\begin{equation}
f_{\rm esc}({\rm LyC}) =\frac{I_{\rm esc}(840)}{I_{\rm mod}(840)}. \label{eq:fesc}
\end{equation}

Using Eq. \ref{eq:fesc} we derive the very high escape fraction 
$f_{\rm esc}$(LyC) = 46$\pm$2 per cent accounting for the uncertainty of the
observed monochromatic flux density.

For the second method \citet{I16b} considered the relation between 
the modelled flux densities of the H$\beta$ emission line $I$(H$\beta$) and 
the intrinsic monochromatic flux density $I_{\rm mod}(900)$. The advantage
of the second method is that there is no need to fit the SED.
The $I$(H$\beta$) is proportional to 
[1-$f_{\rm esc}$(LyC)]$N$(LyC), where $N$(LyC) is the production rate of ionizing
photons in the entire LyC wavelength range. The shape of the ionizing spectrum 
depends on starburst age. Therefore, the $I_{\rm mod}(900)$/$N$(LyC) ratio and 
correspondingly the $I$(H$\beta$)/$I_{\rm mod}(900)$ ratio are not constant and
also depend on starburst age. 
However, \citet{I16b} have shown that, in the case of a small $f_{\rm esc}$(LyC),
the latter ratio can be fairly accurately constrained using a relation between 
the $I$(H$\beta$)/$I_{\rm mod}$(900) ratio and EW(H$\beta$), 
since both the H$\beta$ flux density and the H$\beta$ equivalent width
can directly be derived from observations. 
We have modified eq.~4 in \citet{I16b}
for the case of a large $f_{\rm esc}$(LyC), adopting $\lambda$ = 840\AA\
instead of 900\AA\ and {\sc starburst99} instantaneous burst models 
\citep{L99} for the J1154$+$2443 metallicity:
\begin{equation}
A=\frac{I({\rm H}\beta)}{I_{\rm mod}(840)} = 
3.47\times \left[\frac{{\rm EW}({\rm H}\beta)}{1-f_{\rm esc}({\rm LyC})}\right]^{0.181} \,\,\,\,{\rm \AA}, \label{eq:i900}
\end{equation}
where EW(H$\beta$) is the rest-frame H$\beta$ equivalent width in \AA. 
We finally obtain
\begin{equation}
f_{\rm esc}({\rm LyC})=\frac{I_{\rm esc}(840)}{I_{\rm mod}(840)} =A\frac{I_{\rm esc}(840)}{I({\rm H}\beta)}. \label{eq:fesc1}
\end{equation}

The escape fraction $f_{\rm esc}$(LyC) obtained with the second method, 
derived iteratively from 
Eqs. \ref{eq:i900} and \ref{eq:fesc1} with adopted $I_{\rm esc}$(840) = 
$4.4 \times 10^{-17}$ erg~s$^{-1}$~cm$^{-2}$~\AA$^{-1}$  and $I$(H$\beta$) = 
$9.1 \times 10^{-16}$ erg~s$^{-1}$~cm$^{-2}$ is 47 per cent, in excellent 
agreement with the first method. As in the case of the first method,
only a few iterations are needed for  
$f_{\rm esc}$(LyC) to converge in the second method.

  Thus, we find that the LyC escape fraction in J1154$+$2443 with 
O$_{32}$ = 11.5 is considerably higher than the escape fractions in all known 
low-redshift LyC leakers with lower O$_{32}$ \citep{L13,B14,L16,I16,I16b}.

\subsection{Dependences of $f_{\rm esc}$(LyC) on O$_{32}$ and $V_{\rm sep}$}

In Fig. \ref{fig6}a we present the relation between $f_{\rm esc}$(LyC) and
O$_{32}$ for known low-redshift LyC leaking galaxies. It shows a relatively
tight trend of increasing $f_{\rm esc}$(LyC) with increasing O$_{32}$, 
described by the regression relation
\begin{equation}
f_{\rm esc}({\rm LyC}) = 4.88\times 10^{-3}{\rm O}{_{32}}^2 - 
1.67\times 10^{-2}{\rm O}{_{32}} + 1.07\times 10^{-2}. \label{eq:fesco32}
\end{equation} 
The relation described by Eq. \ref{eq:fesco32} shows that compact low-mass SFGs with high 
O$_{32}$ ratios may lose a considerable fraction of their LyC emission to the 
IGM. 
Thus, we confirm and extend the correlation between 
$f_{\rm esc}$(LyC) and O$_{32}$ first found by \citet{I16b} and later considered
by \citet{F16}.

For comparison, currently the most robust high-redshift Lyman continuum 
leaking galaxy is {\em Ion2} at $z=3.2$ with a relative escape fraction
of 0.64$^{+1.1}_{-0.1}$ \citep{Va15,B16}. This galaxy 
shares many properties with the low-redshift LyC leakers \citep[cf.\ ][]{S16}. 
In particular, {\em Ion2} is characterized by a high ratio 
O$_{32}$ $> 10$, in line with the selection criterion used in this paper
and by \citet{I16,I16b}.

In Fig. \ref{fig6}b we show the tight dependence of $f_{\rm esc}$(LyC)
on the separation $V_{\rm sep}$ between the blue and red peaks of the Ly$\alpha$
emission line in LyC leakers. The regression line
to this dependence is
\begin{equation}
f_{\rm esc}({\rm LyC}) = \frac{4.28\times 10^4}{V{_{\rm sep}}^2} - 
\frac{1.59\times 10^2}{V{_{\rm sep}}} + 0.15, \label{eq:fescVsep}
\end{equation}
where $V_{\rm sep}$ is expressed in km s$^{-1}$. 

We note that the correlation between $f_{\rm esc}$(LyC) and $V_{\rm sep}$
(Fig. \ref{fig6}b) is tighter than that between 
$f_{\rm esc}$(LyC) and O$_{32}$. Physically this may be because both LyC 
and Ly$\alpha$ escaping radiation are determined by the
column density of the neutral gas in LyC leaking galaxies and by the velocity
field in the case of Ly$\alpha$ radiation \citep[e.g. ][]{V15,V17}.
On the other hand, O$_{32}$ depends on several other parameters such as
metallicity and ionization parameter,
in addition to the column density of the neutral gas \citep{JO13,NO14,S15,I17}. 
The relation described by Eq. \ref{eq:fescVsep} might be
important for the construction of radiative transfer models which simultaneously
reproduce $f_{\rm esc}$(LyC) and the Ly$\alpha$ profile.
Similar relations for a smaller sample, with a lower range of 
$f_{\rm esc}$(LyC) and $V_{\rm sep}$, was discussed by \citet{V17}.
The high $f_{\rm esc}$(LyC) and low $V_{\rm sep}$ for J1154$+$2443 are indicative of
a very low neutral gas column density $N$(H {\sc i}). Indeed, according
to radiation-transfer modelling by \citet{V15}, $V_{\rm sep}$ = 199 km s$^{-1}$
would correspond to $N$(H {\sc i}) $\la$ 10$^{18}$ cm$^{-2}$, as predicted by
some models, enabling an efficient leakage of LyC photons.

\section{Conclusions}\label{summary}

In this paper we present new {\sl Hubble Space Telescope} ({\sl HST}) Cosmic
Origins Spectrograph (COS) observations of the 
compact star-forming galaxy (SFG) J1154$+$2443 with a high O$_{32}$ = 
[O~{\sc iii}]$\lambda$5007/[O~{\sc ii}]$\lambda$3727 flux ratio 
$\sim$ 11.5, and a redshift $z$ = 0.3690, aimed at studying its Ly$\alpha$ 
emission and escaping Lyman continuum (LyC) radiation. This study is a 
continuation of a project started by \citet{I16,I16b}.
Our main results are as follows:

1. Escaping LyC radiation is detected in J1154$+$2443 with the highest
escape fraction $f_{\rm esc}$(LyC) = 46~$\pm$~2 per cent found so far in a
low-redshift SFG.

2. A double-peaked Ly$\alpha$ emission line is observed in the
spectrum of J1154$+$2443 with the separation 
between the blue and red peaks $V_{\rm sep}$ of 199 km s$^{-1}$, the lowest 
known so 
far for Ly$\alpha$-emitting galaxies. Comparing the extinction-corrected
Ly$\alpha$/H$\beta$ flux ratio with the case B value, we obtain an escape fraction 
$f_{\rm esc}$(Ly$\alpha$) $\sim$ 98 per cent, among the highest known 
for LAEs. Following \citet{V17} we also find a strong anticorrellation between
the $f_{\rm esc}$ and $V_{\rm sep}$.

3. The COS near ultraviolet (NUV) acquisition images reveal a bright star-forming 
region in the 
centre of the galaxy and an exponential disc at the outskirts, with a disc scale 
length of $\sim$ 1.09 kpc, indicating that J1154$+$2443 is a
dwarf disc system. 

4. J1154$+$2443 is characterized by a high star-formation rate
SFR = 18.9~$M_\odot$~yr$^{-1}$ and a stellar mass $M_\star$ 
$= 10^{8.20}$~M$_\odot$. Its specific star formation rate of 
1.2$\times$10$^{-7}$ yr$^{-1}$ is one of the highest among the known nearby LyC 
leakers.
The metallicity of J1154$+$2443, accurately determined from the optical emission
lines is $12+\log({\rm O/H})=7.65$, the lowest among the known low-redshift
LyC leakers.

5. The observations demonstrate that a selection for compact high-excitation
star-forming galaxies showing a high O$_{32}$ ratio picks up very efficiently 
sources with significant Lyman continuum emission. 
Together with results by \citet{I16,I16b}, our results open 
effective ways to find and explore the sources of cosmic reionization.

\section*{Acknowledgements}

Based on observations made with the NASA/ESA {\sl Hubble Space Telescope}, 
obtained from the data archive at the Space Telescope Science Institute. 
STScI is operated by the Association of Universities for Research in Astronomy,
Inc. under NASA contract NAS 5-26555. Support for this work was provided by 
NASA through grant number HST-GO-14635.001-A from the Space Telescope Science 
Institute, which is operated by AURA, Inc., under NASA contract NAS 5-26555. 
I.O. acknowledges grants GACR 14-20666P and 17-06217Y of the Czech National Foundation. 
Funding for SDSS-III has been provided by the Alfred P. Sloan Foundation, 
the Participating Institutions, the National Science Foundation, and the U.S. 
Department of Energy Office of Science. The SDSS-III web site is 
http://www.sdss3.org/. SDSS-III is managed by the Astrophysical Research 
Consortium for the Participating Institutions of the SDSS-III Collaboration. 
GALEX is a NASA mission  managed  by  the  Jet  Propulsion  Laboratory.
This research has made use of the NASA/IPAC Extragalactic Database (NED) which 
is operated by the Jet  Propulsion  Laboratory,  California  Institute  of  
Technology,  under  contract with the National Aeronautics and Space 
Administration.






\appendix

\section{COS Background Subtraction}
\label{sect:backgroundappendix}

\begin{figure*}
\includegraphics[width=0.7\linewidth]{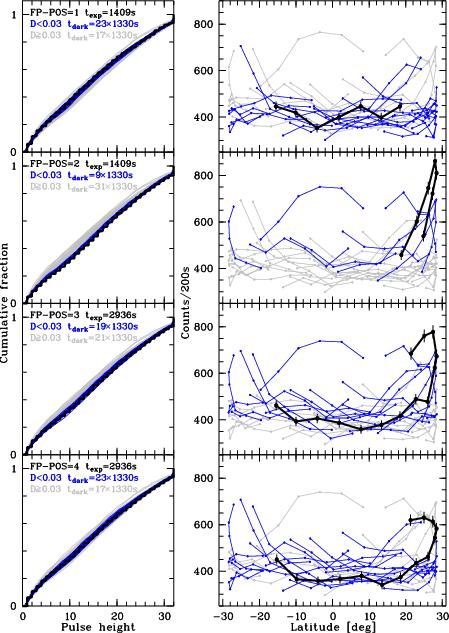}
\caption{\label{figA1}\emph{Left: }Cumulative pulse height distributions in the chosen
calibration windows on COS detector segment A for the four G140L science exposures of J1154$+$2443
(thick black) and for 40 dark exposures obtained within $\pm 1$ month of the science observation date.
Blue lines show dark exposures with a pulse height distribution similar to the one of the science exposure
($D<0.03$), whereas grey lines show rejected dark exposures ($D\ge 0.03$).
\emph{Right: } Corresponding total counts in the calibration windows measured in regular 200\,s time
intervals of the exposure as a function of latitude in {\sl HST}'s orbit. For J1154$+$2443 we also plot
Poisson errors of the counts.}
\end{figure*}

As detailed in Appendix~B of \citet{W16}, the COS background is composed of dark current,
scattered light (mostly scattered geocoronal emission), and the diffuse UV sky background.
At the sensitivity limit of {\sl HST}/COS, the accurary of our flux measurements may be limited
by the accuracy of the estimated total background counts
$B=B_\mathrm{dark}+B_\mathrm{scatter}+B_\mathrm{sky}$ during the exposure.
The dark current $B_\mathrm{dark}$ varies with time, mostly due to variations of the
thermospheric density at {\sl HST}'s altitude with the solar cycle, but also due to changes in the
cosmic ray flux with the Earth's magnetic field in {\sl HST}'s orbit. \citet{W16} found that the
COS detector pulse height distribution of the dark current is sensitive to the ambient conditions,
and that it traces the spatial structure of the dark current across the detector.
Thus, while COS detector gain sag prevents a direct estimation of the dark current in the science
extraction aperture, it can be modelled in post-processing by matching the pulse height distributions
in unilluminated detector regions of the science exposures to those of contemporary dark current monitoring exposures.


\begin{table*}
\caption{\label{tabA1}Counts and estimated backgrounds in the extraction aperture in the rest-frame wavelength range 830--850\,\AA}
\begin{tabular}{lcccccc}\hline
No. of exposure	&$t_\mathrm{exp}$ [s]	&$N$$^{\rm a}$	&$B_\mathrm{dark}$$^{\rm b}$&$B_\mathrm{scatter}$$^{\rm c}$	&$B_\mathrm{sky}$$^{\rm d}$	&$B$$^{\rm e}$
\\ \hline
1		&$1409.184$			&$\cdots$	&$\cdots$				&$\cdots$				&$\cdots$			&$\cdots$\\
2		&$1409.184$			&$58$	&$10.30\pm 0.77$		&$2.08\pm 0.25$		&$0.48\pm 0.05$	&$12.86\pm 1.07$\\
3		&$2936.160$			&$200$	&$29.86\pm 1.61$		&$16.32\pm 1.96$		&$1.67\pm 0.17$	&$47.85\pm 3.74$\\
4		&$2936.192$			&$213$	&$27.76\pm 1.47$		&$16.57\pm 1.99$		&$1.75\pm 0.18$	&$46.08\pm 3.64$\\\hline
total		&$8690.712$			&$471$	&$67.92\pm 3.85$		&$34.97\pm 4.20$		&$3.89\pm 0.39$	&$106.78\pm 8.44$\\
\hline
\end{tabular}

\hbox{$^{\rm a}$Total number of counts (object+background).}

\hbox{$^{\rm b}$Counts of the dark current.}

\hbox{$^{\rm c}$Counts of the scattered light (mostly scattered geocoronal 
emission).}

\hbox{$^{\rm d}$Counts of the diffuse UV sky background.}

\hbox{$^{\rm e}$Total number of the background counts.}

\end{table*}

\begin{figure*}
\includegraphics[width=0.7\linewidth]{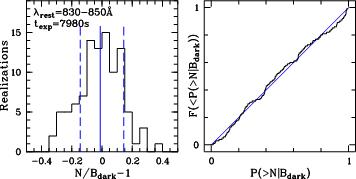}
\caption{\label{figA2}\emph{Left: }Relative deviation between measured dark counts $N$
and estimated dark current $B_\mathrm{dark}$ for random subsamples of 6/60 dark exposures obtained
2017 June 5 -- August 21 in the aperture and wavelength range used to measure the LyC flux.
The solid and dashed lines show the mean and the standard deviation of the 100 realizations, respectively.
\emph{Right: }Cumulative fraction $F\left(<P\left(>N|B_\mathrm{dark}\right)\right)$ of Poisson probabilities
$P\left(>N|B_\mathrm{dark}\right)$ to detect more than $N$ dark counts given the dark current
$B_\mathrm{dark}$ for the 100 realizations. The blue line marks identity.}
\end{figure*}

Specifically, for our COS G140L data obtained at COS Lifetime Position 3 we chose two 50-pixel wide
unilluminated regions above and below the trace (geometrically corrected spatial coordinates $379<y<429$ and $502<y<552$),
avoiding spectral ranges affected by scattered geocoronal emission, grid wires and detector blemishes.
In the left panels of Fig.~\ref{figA1} we compare the cumulative pulse height distributions
in the calibration regions of the four science exposures to those same regions in 40 dark monitoring exposures
(Programme 14520) obtained within $\pm 1$ month of the science observations. We characterised their
difference by computing the maximum absolute difference in the cumulative distributions
\begin{equation}
D=\max_k\left|F_\mathrm{dark}\left(<k\right)-F_\mathrm{science}\left(<k\right)\right|\,, 
\end{equation}
where $F_\mathrm{dark}$ and $F_\mathrm{science}$ are the empirical cumulative distributions to
pulse height amplitude $0\le k\le 31$. Due to low solar activity in the sampled time interval the pulse height
distributions are quite similar, with remaining small differences due to varying local ambient conditions. 
We chose a threshold $D<0.03$ to combine the dark exposures obtained in similar conditions to the science
exposures, treating science exposures individually. The dark current was extracted in the chosen science
aperture with the same pulse height thresholds as for the science data, and smoothed with a 500-pixel
running average to capture variations in the dark current along the dispersion axis, avoiding grid wires and
detector blemishes. Finally, the smoothed dark current was scaled to the science exposure via the ratio
of the total counts in the calibration regions. The dark current was co-added together with the science exposures,
accounting for focal plane offsets and propagating statistical errors of the scaling factors and Poisson errors
in the smoothed darks. The final dark current estimate is accurate to a few per cent.

The right panels in Fig.~\ref{figA1} show the total counts in the calibration regions obtained in
regular 200\,s intervals during the exposure as a function of geographic latitude of {\sl HST} in its orbit.
The dark current increases from low to high geographic latitude by $\sim 50$ per cent, but with statistically
significant scatter even at the well-sampled low latitudes. The increase at high geographic latitudes is most
likely due to the increased cosmic ray flux at high geomagnetic latitudes 
\citep[see e.g.\ ][ for
measurements with {\sl HST}/GHRS]{H95}. The scatter at fixed geographic latitude is due to the tilt between the
geomagnetic axis and Earth's rotation axis, but also due to longitudinal features in the Earth's magnetic field,
such as the South Atlantic Anomaly. Thus, part of the variation in the average dark count rate measured in
the dark monitoring programme is due to different orbit parameters and locations sampled by the observations,
in addition to the much larger variation during solar activity (low in the time interval sampled in Fig.~\ref{figA1}).
Due to the sparse dark current monitoring ($5\times 1330$\,s per week) randomly sampling the part of the
magnetosphere traversed by {\sl HST} in its orbit, it is likely impossible to obtain a sufficiently accurate map of the
dark current as a function of geomagnetic coordinates, even when averaging over long time intervals during solar minimum.
Therefore, we approximated the dark current during the science observations with dark monitoring data
obtained at vastly different locations but with a similar pulse height distribution. Accounting for the difference in
exposure times (i.e.\ the range in latitude) Fig.~\ref{figA1} indicates that our threshold $D<0.03$
selects darks obtained at similar latitude as the science observations.

We tested our approach by treating random subsets of 6 out of 60 dark exposures obtained between June and August 2017
as science data, with the remainder taken for calibration \citep[e.g.][]{W16}. Focal plane offsets were emulated
(1330\,s at offset positions 1 and 2, $2\times 1330$\,s at positions 3 and 4), such that the co-added spectrum
had a pixel exposure time distribution very similar to the G140L data of J1154$+$2443.
We used the same calibration regions and pulse height range as in the science reductions.
The left panel of Fig.~\ref{figA2} shows the relative deviation between the measured dark counts
and the estimated dark current in the aperture and wavelength range of our LyC flux measurement in J1154$+$2443.
From 100 realizations we obtain a mean relative deviation $-0.012$, showing that our procedure estimates
the dark current with insignificant systematic error \citep[see also][]{W16,C17}. The rather large standard deviation
of $0.144$ is due to Poisson fluctuations of the measured dark counts around the mean $B_\mathrm{dark}\simeq 50$
counts\footnote{The estimated $B_\mathrm{dark}$ varies depending on the chosen subset of darks treated as science exposures.}.
To test this we computed the Poisson probability
\begin{equation}\label{eq:prob}
P\left(>N|B_\mathrm{dark}\right)=1-\sum_{k=0}^{N}\frac{B_\mathrm{dark}^k e^{-B_\mathrm{dark}}}{k!}
\end{equation}
of measuring more than the $N$ registered counts given the estimated dark current $B_\mathrm{dark}$,
and compared that number to the fraction of the 100 realizations with smaller $P$ values \citep{W16}.
The right panel of Fig.~\ref{figA2} shows that the empirical cumulative distribution approximately
follows the expectation for a Poisson distribution, in particular in the tails of strong over- and undersubtraction
of the dark current. This analysis also confirms that the dark current of a particular exposure can be modelled
satisfactorily with dark exposures obtained at different locations.

As a result, Equation~\ref{eq:prob} estimates the probability that the measured counts $N>B$ are consistent
with a Poisson fluctuation of the background $B$. The scattered light and its uncertainty was estimated following
\citet{W16}, whereas the diffuse sky background was taken from \citet{M14} with an estimated uncertainty of 10 per cent.
Table~\ref{tabA1} lists the total counts and the estimated background components in the rest-frame
wavelength range used for our LyC flux measurement in J1154$+$2443. Note that due to the focal plane offsets
and the grid wires the actual spectral coverage differs among the four exposures. The detected counts significantly
exceed the total estimated background in every exposure, and the estimated background uncertainty does not
affect this result. In the co-added exposure, it is highly unlikely that the measured 471 counts in the LyC are entirely
due to the total background ($P\left(>N|B\right)<10^{-7}$ eventually limited by the assumption in Equation~\ref{eq:prob}).
Even assuming a background higher by a factor $3.5$ would not change this result ($P\left(>N|B\right)=5.4\times 10^{-7}$).

\bsp	
\label{lastpage}
\end{document}